\pdfoutput=1
\documentclass[iop]{emulateapj}
\usepackage{amsmath,amssymb,color,verbatim}
\slugcomment{}

\shorttitle{Investigating the physical conditions in extended system hosting mid-infrared bubble N14}
\shortauthors{L.~K. Dewangan et al.}
\begin{document}
\title{Investigating the physical conditions in extended system hosting mid-infrared bubble N14}
\author{L.~K. Dewangan\altaffilmark{1}, T. Baug\altaffilmark{2}, L.~E. Pirogov\altaffilmark{3}, and D.~K. Ojha\altaffilmark{4}}
\email{lokeshd@prl.res.in}
\altaffiltext{1}{Physical Research Laboratory, Navrangpura, Ahmedabad - 380 009, India.}
\altaffiltext{2}{Kavli Institute for Astronomy and Astrophysics, Peking University, 5 Yiheyuan Road, Haidian District, Beijing 100871, P. R. China.}
\altaffiltext{3}{Institute of Applied Physics of the Russian Academy of Sciences, 46 Ulyanov st., Nizhny Novgorod 603950, Russia.}
\altaffiltext{4}{Department of Astronomy and Astrophysics, Tata Institute of Fundamental Research, Homi Bhabha Road, Mumbai 400 005, India.}
\begin{abstract}
To observationally explore physical processes, we present a multi-wavelength study of a wide-scale 
environment toward {\it l} = 13$\degr$.7--14$\degr$.9 containing a mid-infrared bubble N14. 
The analysis of $^{12}$CO, $^{13}$CO, 
and C$^{18}$O gas at [31.6, 46] km s$^{-1}$ reveals an extended physical system (extension $\sim$59 pc $\times$ 29 pc), which hosts at least five groups of the ATLASGAL 870 $\mu$m dust clumps at d $\sim$3.1 kpc. These spatially-distinct groups/sub-regions 
contain unstable molecular clumps, and are associated with several Class~I young stellar objects (mean age $\sim$0.44 Myr). 
At least three groups of ATLASGAL clumps 
associated with the expanding H\,{\sc ii} regions (including the bubble N14) and 
embedded infrared dark clouds, devoid of the ionized gas, are found in the system. 
The observed spectral indices derived using the GMRT and THOR radio continuum 
data suggest the presence of non-thermal emission with the H\,{\sc ii} regions.
High resolution GMRT radio continuum map at 1280 MHz traces several ionized clumps 
powered by massive B-type stars toward N14, which are considerably young (age $\sim$10$^{3}$--10$^{4}$ years). Locally, early stage of star formation is evident toward all the groups of clumps. 
The position-velocity maps of $^{12}$CO, $^{13}$CO, 
and C$^{18}$O exhibit an oscillatory-like velocity pattern toward the selected longitude range. 
Considering the presence of different groups/sub-regions in the system, the oscillatory pattern in velocity is indicative of the 
fragmentation process. All these observed findings favour the applicability of the global collapse scenario in the extended physical system, which also seems to explain the observed hierarchy.
\end{abstract}
\section{Introduction}
\label{sec:intro}
It has been well accepted that molecular gas converts into young stellar objects (YSOs) of different masses (including massive OB stars ($\gtrsim$ 8 M$_{\odot}$)) and their clusters in a giant molecular cloud (GMC), where several complex physical processes may operate. However, the mechanisms responsible for the birth of young stellar clusters and massive stars are still incompletely understood \citep{zinnecker07,tan14}. The study of a GMC allows to explore the ongoing star-formation mechanisms, such as global gravitational contraction \citep[][and references therein]{hartmann12}, triggered star formation scenarios \citep[i.e., ``globule squeezing", ``collect and collapse", and ``cloud-cloud collision (CCC)";][]{elmegreen98}. 
Such study requires the knowledge of the physical conditions in promising massive star-forming sites (e.g., H\,{\sc ii} regions) associated with a GMC, which can be inferred through the analysis of the multi-wavelength data. In this context, the present paper deals with an extended and a single physical system hosting several massive star-forming regions, which are 
situated toward {\it l} = 13$\degr$.7--14$\degr$.9 and {\it b} = $-$0$\degr$.5--+0$\degr$.1. This extended physical system located in the Galactic arm(s) is identified via the reliable information of distance and radial velocity ($V_\mathrm{lsr}$) of its different sub-regions, which helps to disentangle the target system against its background and foreground clouds. 

Figure~\ref{fig1}a shows the APEX Telescope Large Area Survey of the Galaxy \citep[ATLASGAL; beam size $\sim$19$\farcs$2;][]{schuller09} 870 $\mu$m dust continuum map 
covering the wide-scale area ($\sim$1$\degr$.2 $\times$ 0$\degr$.6) around the mid-infrared (MIR) bubble N14. The ATLASGAL map is also overlaid with 53 ATLASGAL 
clumps \citep[taken from][]{urquhart18}. All these clumps are traced in a velocity range of [34.5, 43] km s$^{-1}$, and are located at a single distance of $\sim$3.1 kpc \citep{urquhart18}. 
The spatial distribution of these clumps has enabled us to find out an extended physical system. Based on the visual inspection, one can also arbitrarily depict at least five groups of clumps (i.e., group1, group2, group3, group4, and group5) in Figure~\ref{fig1}a, which are indicated by broken curves. The MIR bubble N14 has been characterized as a complete or closed ring with an average radius and thickness of 1$'$.22 and 0$'$.38, respectively \citep{churchwell06,dewangan13,yan16}. 
The bubble N14 also contains the ionized emission at its center \citep[e.g.,][]{dewangan13}. 
Using the Multi-Array Galactic Plane Imaging Survey \citep[MAGPIS; beam size $\sim$6$''$;][]{helfand06} radio continuum flux at 20~cm of the bubble N14, \citet{beaumont10} computed the Lyman continuum photons ($\log{N_\mathrm{uv}}$) to be 48.36 \citep[see also][]{dewangan13}, which is explained by a single O9V--O8.5V star \citep{panagia73} or a single O7.5V--O8V star \citep{martins05} 
or at least six O9.5V stars \citep{beaumont10}. 
In addition to the bubble N14, some other previously known sources (such as, G014.194$-$00.194, G14.427$-$00.075HII, G14.47$-$0.20, IRAS 18141$-$1615, and G14.71$-$0.19) are also labeled in 
Figure~\ref{fig1}a. 
Figure~\ref{fig1}b displays the overlay of the MAGPIS 20~cm continuum emission contours on the ATLASGAL map at 870~$\mu$m. In the direction of at least three groups of clumps (i.e., group2, group3, and group4 in 
Figure~\ref{fig1}a), the MAGPIS 20~cm contours reveal the presence of H\,{\sc ii} regions powered by massive OB 
stars. 

However, the formation processes of the selected different groups of clumps as well as massive OB stars are yet to be investigated in the extended 
physical system (see Figure~\ref{fig1}a). No attempt is made to examine the velocity structure of molecular gas and the identification of YSOs toward 
the entire selected longitude range. Such analysis is required for exploring the ongoing star formation mechanisms in 
the extended physical system containing several embedded clumps and H\,{\sc ii} regions. It also helps us to observationally understand the origin of the large-scale configuration/system toward {\it l} = 13$\degr$.7--14$\degr$.9. 
In this context, a multi-wavelength approach is adopted in this paper, which is a very useful and effective utility to gain the quantitative and qualitative physical information in the target site. The present work is benefited with the existing large-scale FOREST Unbiased Galactic plane Imaging survey with the Nobeyama 45-m 
telescope \citep[FUGIN;][]{umemoto17} molecular line data (i.e., $^{12}$CO, $^{13}$CO, and C$^{18}$O) along with the {\it Spitzer} and {\it Herschel} infrared 
maps. New radio continuum maps observed by Giant Metrewave Radio Telescope (GMRT) facility are also presented toward the MIR bubble N14. 

This paper is arranged as follows. Section~\ref{sec:obser} presents the details of the adopted data sets in this paper. 
Section~\ref{sec:data} provides new outcomes derived using a multi-wavelength approach in the selected longitude range. 
The discussion of the observational outcomes is presented in Section~\ref{sec:disc}. 
Finally, Section~\ref{sec:conc} gives the summary of the major findings obtained in this work.
\section{Data and analysis}
\label{sec:obser}
The present paper utilizes the existing multi-wavelength data sets obtained from various large-scale surveys (see Table~\ref{xftab1}). 
The selected target area ($\sim$1$\degr$.2 $\times$ 0$\degr$.6 (or $\sim$65 pc $\times$ 32.5 pc); centered at $l$ = 14$\degr$.3; $b$ = $-$0$\degr$.2) in this 
paper is presented in Figure~\ref{fig1}a. 
The $^{12}$CO(J =1$-$0), $^{13}$CO(J =1$-$0), and C$^{18}$O(J =1$-$0) line data were obtained from the FUGIN 
survey, and are calibrated in main beam temperature ($T_\mathrm{mb}$) \citep{umemoto17}. 
The typical rms noise level\footnote[1]{https://nro-fugin.github.io/status/} ($T_\mathrm{mb}$) is $\sim$1.5~K, $\sim$0.7~K, and $\sim$0.7~K for 
$^{12}$CO, $^{13}$CO, and C$^{18}$O lines, respectively \citep{umemoto17}. To improve sensitivities, each FUGIN molecular line data cube is smoothened with a Gaussian function having half power beam width of 35$''$. Additionally, in the direction of the bubble N14, our unpublished GMRT radio continuum maps are also examined in this work. 

Radio continuum observations at 610 and 1280 MHz were performed with the GMRT facility on 2012 December 28 (Proposal Code: 23$\_$054; PI: L.~K. Dewangan). 
The GMRT data were reduced using the Astronomical Image Processing System (AIPS) package, and the detailed reduction procedures can be found in \citet{mallick12,mallick13}. 
We flagged out bad data from the UV data by multiple rounds of flagging using the {\sc tvflg} task of the AIPS. 
After several rounds of `self-calibration', the final maps at 610 and 1280 MHz were produced with the synthesized beams 
of 5$''$.6 $\times$ 5$''$.2 and 6$''$ $\times$ 6$''$, respectively. 
In general, due to the Galactic background emission, the antenna temperature of the sources located toward the Galactic plane is expected to be increased. It is found more severe in the GMRT low frequency bands (i.e., 610 MHz), 
where the contribution from the background emission is generally much higher compared to the 1280 MHz band. 
More detailed description concerning the system temperature correction can be found in \citet[][and references therein]{baug15}. 
Such correction is also applied to the GMRT 610 MHz data, before performing any scientific analysis. 
The final rms sensitivities of both the maps at 610 and 1280 MHz are $\sim$1 mJy beam$^{-1}$. The unit of brightness in Jy~beam$^{-1}$ is adopted in this paper. 
However, the conversion from Jy~beam$^{-1}$ to Jy sr$^{-1}$ is Jy~beam$^{-1}\times(\frac{\theta}{206265})^{2}\times\frac{\pi}{4\ln{2}}$= Jy sr$^{-1}$, where $\theta$ is the 
beam size in arcseconds.
\begin{table*}
\setlength{\tabcolsep}{0.0in}
\centering
\caption{Table provides the list of different surveys adopted in this paper.}
\label{xftab1}
\begin{tabular}{lcccr}
\hline 
  Survey  & band/line(s)       &  Resolution ($\arcsec$)        &  Reference \\   
\hline
\hline 
 Multi-Array Galactic Plane Imaging Survey (MAGPIS)                             & 20 cm                       & $\sim$6          & \citet{helfand06}\\
 The HI/OH/Recombination line survey of the inner Milky Way (THOR)                             & 1--2 GHz                       & $\sim$25          & \citet{beuther16}\\
GMRT observations (Proposal Code: 23$\_$054)  &   610 MHz, 1280 MHz &$\sim$5--6     &  PI: L.~K. Dewangan\\
FUGIN survey  &  $^{12}$CO, $^{13}$CO, C$^{18}$O (J = 1--0) & $\sim$20        &\citet{umemoto17}\\
APEX Telescope Large Area Survey of the Galaxy (ATLASGAL)                 &870 $\mu$m                     & $\sim$19.2        &\citet{schuller09}\\
{\it Herschel} Infrared Galactic Plane Survey (Hi-GAL)                              &70--500 $\mu$m                     & $\sim$5.8--37         &\citet{molinari10}\\
{\it Spitzer} MIPS Inner Galactic Plane Survey (MIPSGAL)                                         &24 $\mu$m                     & $\sim$6         &\citet{carey05}\\ 
{\it Spitzer} Galactic Legacy Infrared Mid-Plane Survey Extraordinaire (GLIMPSE)       &3.6--8.0  $\mu$m                   & $\sim$2           &\citet{benjamin03}\\
\hline          
\end{tabular}
\end{table*}
\section{Results}
\label{sec:data}
In the selected target area, the multi-wavelength data sets are analyzed to study the distribution of molecular gas, YSOs, 
embedded clumps, H\,{\sc ii} regions, dust temperature as well 
as velocity structure.
\subsection{Extended physical system hosting H\,{\sc ii} regions}
\label{subsec:radio1}
\begin{figure*}
\epsscale{0.95}
\plotone{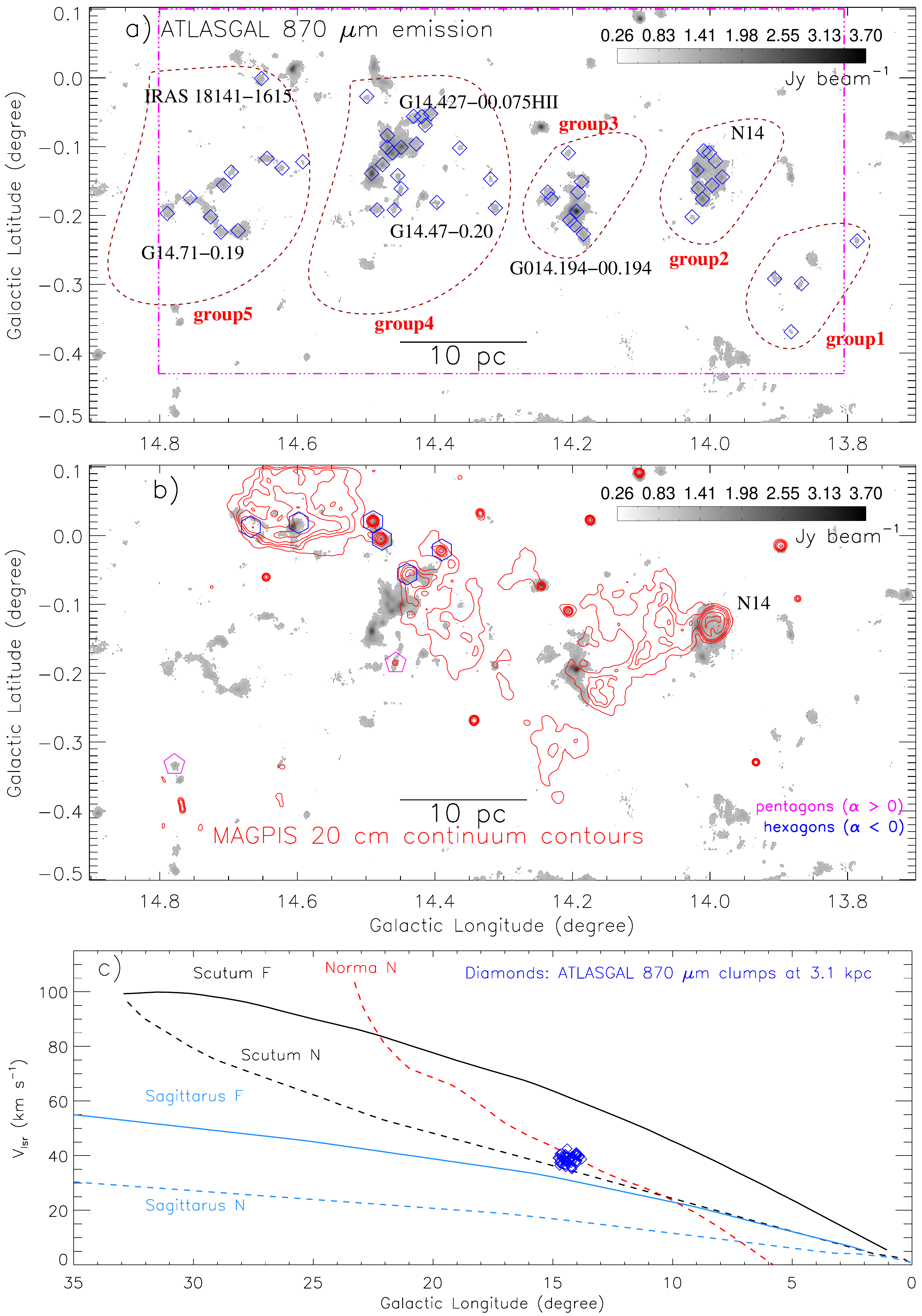}
\caption{a) The panel displays the ATLASGAL contour map at 870 $\mu$m (area $\sim$1$\degr$.2 $\times$ 0$\degr$.6 
($\sim$65 pc $\times$ 32.5 pc at a distance of 3.1 kpc); central coordinates: $l$ = 14$\degr$.3; $b$ = $-$0$\degr$.2) overlaid
 with the ATLASGAL dust continuum clumps at 870 $\mu$m \citep[from][]{urquhart18} (see diamonds). 
The ATLASGAL contours are shown with the levels of 3.9 Jy beam$^{-1}$ $\times$ (0.067, 0.08, 0.1, 0.15, 0.2, 0.25, 0.3, 0.4, 
0.5, 0.6, 0.7, 0.8, 0.9, and 0.95). 
All the ATLASGAL clumps are located at a distance of $\sim$3.1 kpc. 
The broken box (in magenta) refers to the area shown using the MAGPIS 20 cm continuum data in Figure~\ref{fig1}b. 
Some known regions (e.g., bubble N14, G014.194-00.194, G14.427-00.075HII, G14.47-0.20, IRAS 18141-1615, and G14.71-0.19) 
are also labeled in the figure. At least five groups of ATLASGAL clumps are indicated in the figure (see broken curves).
b) Overlay of the MAGPIS 20 cm continuum contours (in red; resolution $\sim$6$''$) on the ATLASGAL contour map at 870 $\mu$m. 
The ATLASGAL map is the same as in Figure~\ref{fig1}a. The MAGPIS contours are shown with the levels 
of 2.2, 2.8, 3.3, 4.0, 5.5, and 8.0 mJy beam$^{-1}$. The MAGPIS map is smoothed using a Gaussian function 
with radius of four pixels. Hexagon and Pentagon symbols represent the radio continuum sources from the 
THOR survey \citep[{\it l} $>$ 14$\degr$.3;][]{bihr16,wang18}. 
Pentagon symbols show the sources with spectral index ($\alpha$) $>$ 0, while the sources with $\alpha$ $<$ 0 are marked by hexagon symbols. 
c) The panel displays the Sagittarius, Scutum, and Norma arms toward {\it l} = 0$\degr$.0 -- 35$\degr$.0 
\citep[from][]{reid16} in longitude-velocity plot. The near and far sides of the arms are presented by broken and solid curves, respectively. 
The $V_\mathrm{lsr}$ of each ATLASGAL clump against its longitude is also marked in the plot. 
In panels a) and b), the scale bar referring to 10 pc (at a distance of 3.1 kpc) is drawn.} 
\label{fig1}
\end{figure*}

In the target longitude range (i.e., {\it l} = 13$\degr$.7--14$\degr$.9), different groups of the ATLASGAL 870 $\mu$m dust 
continuum clumps (at $V_\mathrm{lsr}$ range $\sim$[34.5, 43] km s$^{-1}$) are presented in Section~\ref{sec:intro}, and are labeled in Figure~\ref{fig1}a. In the direction of some of these ATLASGAL groups, the radio continuum emission is observed (see Figure~\ref{fig1}b). 
The dust continuum emission at 870 $\mu$m may depict cold dust, while the ionized gas is traced by the radio continuum emission. 
As highlighted earlier, the extended physical system hosts several embedded 
clumps and H\,{\sc ii} regions powered by massive OB stars. Hence, Figure~\ref{fig1}b helps us to examine the spatial 
association between the dust clumps and the ionized gas in the system.
 
In Figure~\ref{fig1}b, we have marked the positions of the radio continuum sources (for {\it l} $>$ 14$\degr$.3) from 
the THOR survey \citep{bihr16,wang18}, which are shown by hexagons and pentagons. The THOR radio sources are detected toward the ATLASGAL clumps and 
the MAGPIS radio continuum emission. Each THOR radio source has spectral index ($\alpha$) value, which is defined as $F_\nu$ $\propto$ $\nu^{\alpha}$. 
Here, $\nu$ is the frequency of observation, and $F_\nu$ is the corresponding observed flux density. 
The spectral indices of THOR radio sources are derived 
using the radio peak fluxes at 1.06, 1.31, 1.44, 1.69, 1.82, and 1.95 GHz \citep[e.g.,][]{bihr16}. 
Note that the areas around the bubble N14 are not observed in the THOR survey. 
THOR sources with $\alpha$ $<$ 0 are shown by hexagons, while pentagons indicate THOR sources with $\alpha$ $>$ 0. 
These sources are G14.779$-$0.333 ($\alpha$ $\sim$2.25), G14.457$-$0.185 ($\alpha$ $\sim$0.58), G14.477$-$0.005 ($\alpha$ $\sim-0$.097), 
G14.490+0.021 ($\alpha$ $\sim-0$.026), G14.598+0.019 ($\alpha$ $\sim-$0.042), 
G14.668+0.013 ($\alpha$ $\sim-$0.599), G14.390$-$0.021 ($\alpha$ $\sim-$0.074), and G14.440$-$0.056 ($\alpha$ $\sim-$0.344). 
In general, the positive and negative values of $\alpha$ allow us to distinguish 
the thermal and non-thermal radio continuum emission in a given massive star-forming region, respectively.  
A positive or near zero spectral index refers to thermally emitting sources \citep{bihr16}. 
For example, supernova remnants (SNR) display non-thermal emission with $\alpha$ $\approx$ $-$0.5, while a steeper $\alpha$ $\approx$ $-$1 is 
expected in extragalactic objects \citep[e.g.,][]{rybicki79,longair92,bihr16}.  
Based on the radio morphology, all the selected 8 THOR radio sources appear to be Galactic H\,{\sc ii} regions. 
Six out of 8 THOR radio sources show $\alpha$ $<$ 0, suggesting the non-thermal radio continuum emission, 
and the remaining 2 THOR sources exhibit thermal radio continuum emission (or free-free emission).  

In the catalog of ATLASGAL clumps \citep[e.g.,][]{urquhart18}, one can obtain the integrated flux and effective radius ($R_\mathrm{c}$) of 
each ATLASGAL clump as well as other parameters, such as distance, $V_\mathrm{lsr}$, dust temperature ($T_\mathrm{d}$), bolometric luminosity ($L_{\rm bol}$), 
clump mass ($M_\mathrm{clump}$), and H$_{2}$ column density ($N(\mathrm H_2)$). 
Table~\ref{tab3} lists the positions and physical parameters of all these clumps. 
Additionally, we have also included the average volume 
density ($n_{\mathrm H_2}$ = 3$M_\mathrm{clump}$/(4$\pi$$R_\mathrm{clump}^{3}$$\mu_{\rm H_2} m_{\rm H}$)) of each clump in the table. 
In the calculation, we assume that each clump has a spherical geometry. The mean molecular weight $\mu_{\rm H_2}$ is adopted 
to be 2.8, and $m_{\rm H}$ is the mass of an hydrogen atom. 

\begin{figure*}
\epsscale{1}
\plotone{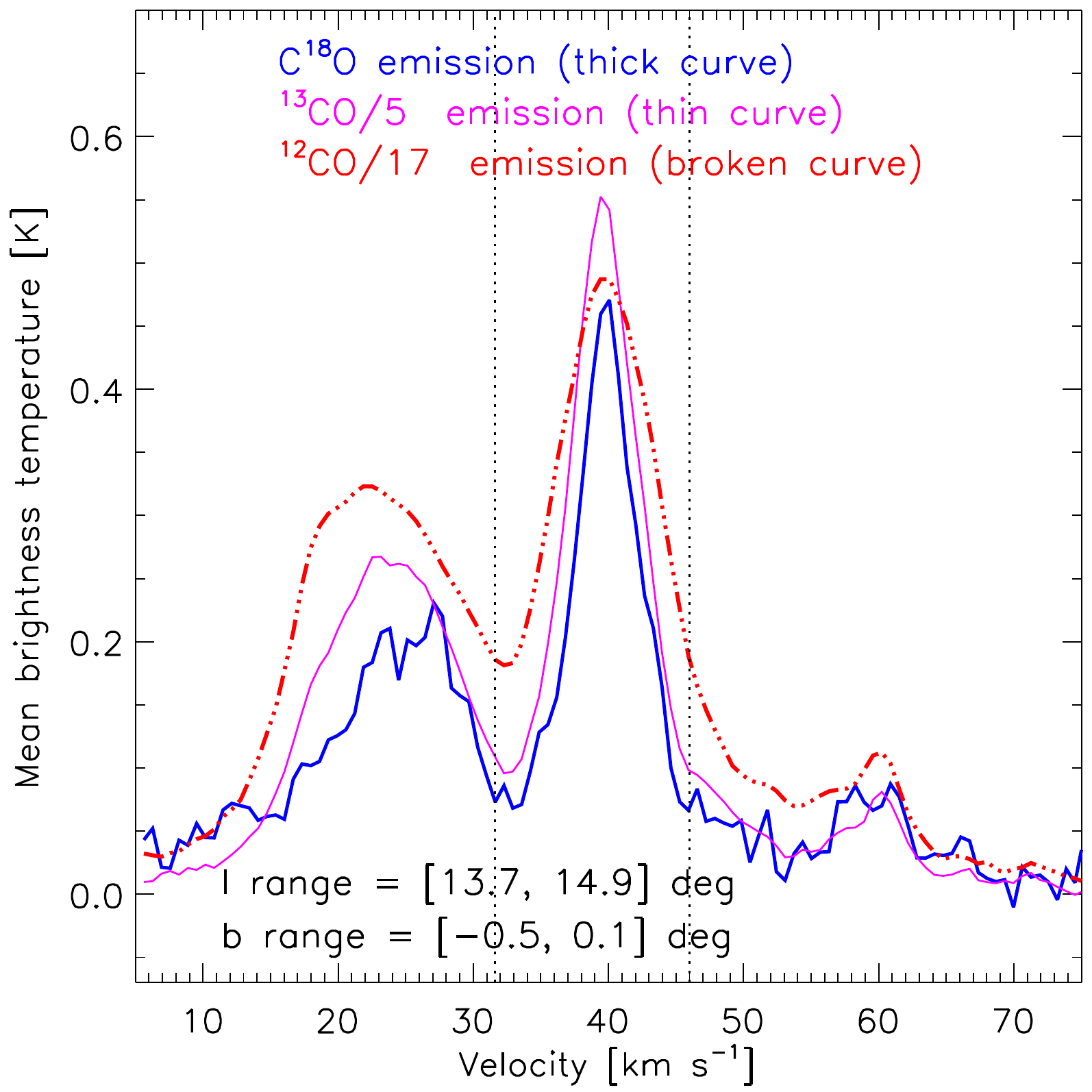}
\caption{The panel shows the FUGIN $^{12}$CO profile (broken red curve), $^{13}$CO spectrum (thin magenta curve), and C$^{18}$O 
profile (thick blue curve). The profiles are produced by averaging the target area shown in 
Figure~\ref{fig1}a. Ranges of the longitude and latitude of the target area are also labeled in the figure. The $^{12}$CO and $^{13}$CO spectra have been divided by a factor of 5 and 17, respectively.} 
\label{zfig1}
\end{figure*}

Figure~\ref{fig1}c presents a plot of $V_\mathrm{lsr}$ of 53 ATLASGAL clumps 
vs. Galactic longitude range of {\it l} = 0$\degr$ -- 35$\degr$. 
The locations of various spiral arms (i.e., near and far sides of the Sagittarius, Scutum, 
and Norma arms) of the Milky Way \citep[from][]{reid16} are also marked in Figure~\ref{fig1}c.
This analysis suggests that the cloud associated with the extended physical system is located toward 
the near sides of the Scutum and Norma arms. 
In Figure~\ref{zfig1}, we present the observed $^{12}$CO, $^{13}$CO, and C$^{18}$O 
spectra toward the selected target area. These profiles are obtained by averaging the selected target area as presented in Figure~\ref{fig1}a. 
In Figure~\ref{zfig1}, we find three velocity peaks around 23, 40, and 60 km s$^{-1}$. 
The extended physical system, containing H\,{\sc ii} regions (including the bubble N14), is associated with 
the velocity component around 40 km s$^{-1}$, and is well depicted in a velocity 
range of [31.6, 46] km s$^{-1}$. 
Note that the observed velocities of all the selected 
ATLASGAL clumps are well fallen within this velocity range. 
This exercise also indicates that the extended physical system is not physically associated with other two velocity 
components around 23 and 60 km s$^{-1}$, which are not examined in this paper.

\begin{figure*}
\epsscale{0.59}
\plotone{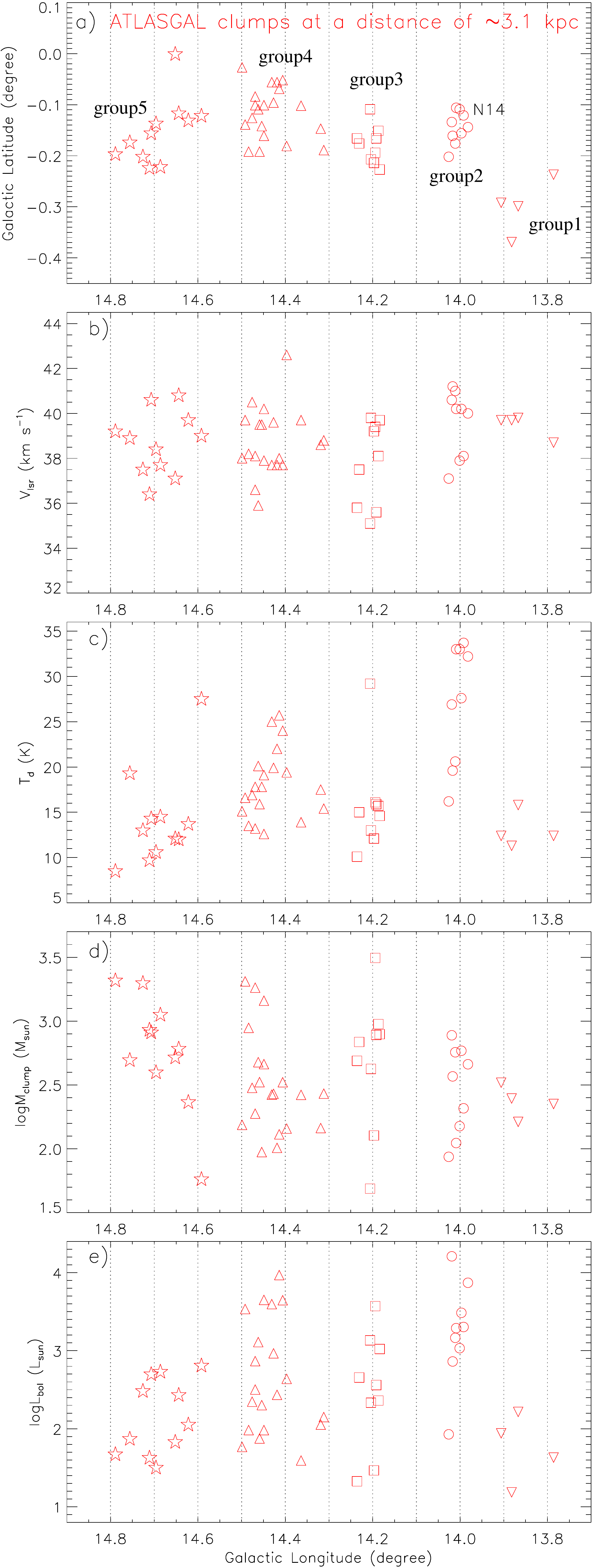}
\caption{a) Spatial distribution of 53 ATLASGAL dust continuum clumps at 870 $\mu$m toward the selected site. 
The ATLASGAL dust continuum clumps at 870 $\mu$m \citep[from][]{urquhart18} are highlighted by different 
symbols (i.e., up down triangles, circles, squares, triangles, and stars), which 
are used to show five groups of clumps (see Table~\ref{tab3}).
These groups are labeled as group1 (up down triangles), group2 (circles), group3 (squares), 
group4 (triangles), and group5 (stars). b-c-d-e) Distribution of the $V_\mathrm{lsr}$, dust temperature, mass, 
and bolometric luminosity of clumps against the Galactic longitude.}
\label{fig2}
\end{figure*}

To display five groups of clumps, Figure~\ref{fig2}a shows the positions of 53 ATLASGAL 
clumps at 870 $\mu$m using different symbols (i.e., up down triangles (group1), circles (group2), squares (group3), 
triangles (group4), and stars (group5)). The sites N14, G014.194$-$00.194, G14.427$-$00.075HII (and G14.47$-$0.20), 
G14.71$-$0.19 are seen toward group2, group3, group4, and group5, respectively. Figure~\ref{fig2}b shows the distribution of $V_\mathrm{lsr}$ of 53 clumps against 
Galactic longitude. 
We find a noticeable velocity spread toward all the ATLASGAL groups (except group1), which is further explored using 
the molecular line data in Section~\ref{sec:coem}. In Figure~\ref{fig2}c, we display the distribution of the dust temperatures of clumps against the Galactic longitude, 
showing a dust temperature range of $\sim$8--34~K. In the group3 and group5, the clumps are found with $T_\mathrm{d}$ $<$ 15~K. 
In the direction of the bubble N14, the clumps associated with the group2 have $T_\mathrm{d}$ $>$ 25~K. 
One can find the masses and bolometric luminosities of the clumps associated with different ATLASGAL groups in 
Figures~\ref{fig2}d and~\ref{fig2}e, respectively. All the ATLASGAL groups (except group1 and group5) have clumps 
with $L_{\rm bol}$ $>$ 10$^{3}$~L$_{\odot}$. 
In each group, at least two dense clumps (with $n_{\mathrm H_2}$ $>$10$^{5}$ cm$^{-3}$) are found (see Table~\ref{tab3}). 
\subsection{Kinematics of molecular gas}
\label{sec:coem} 
\begin{figure*}
\epsscale{0.87}
\plotone{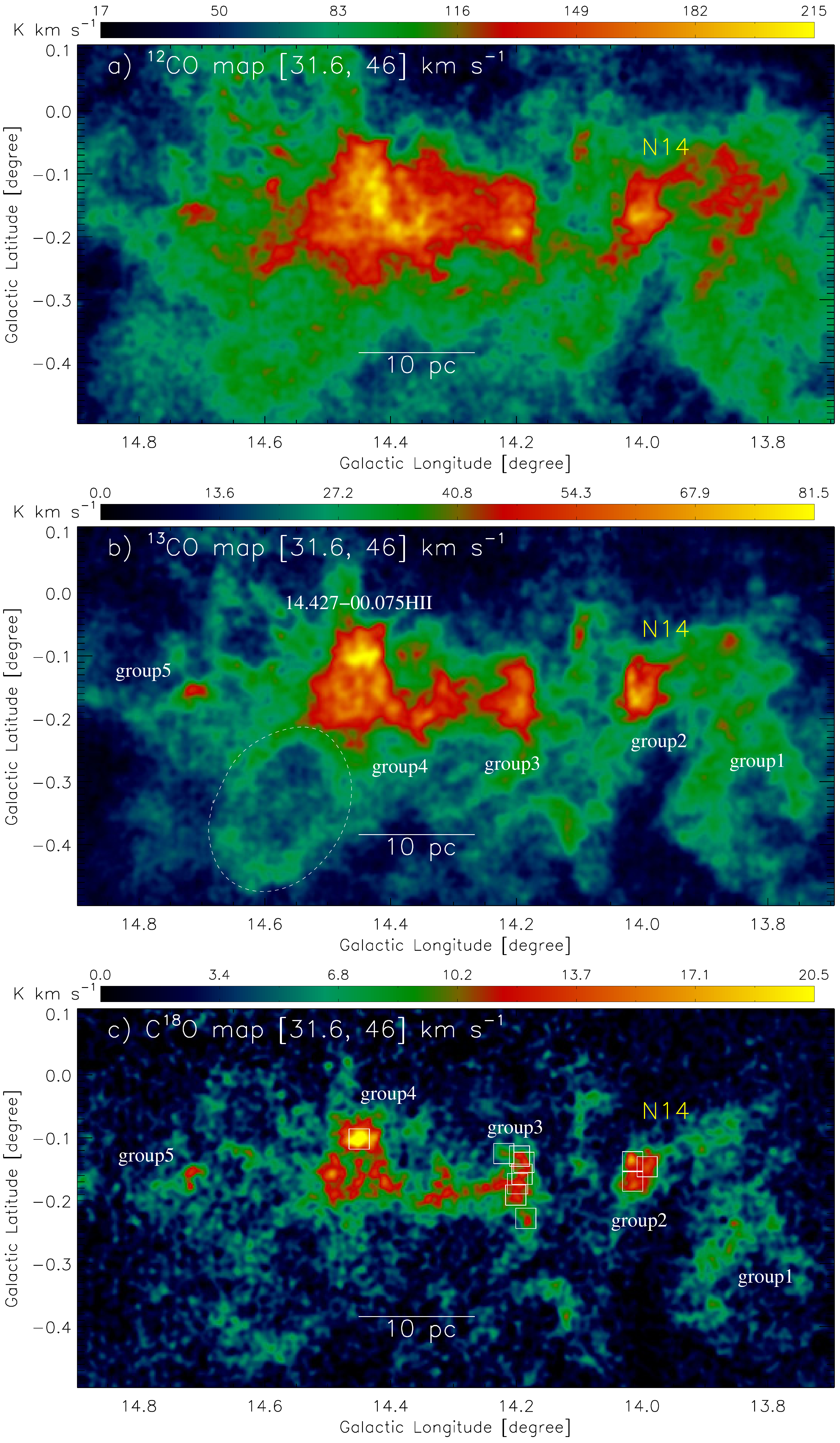}
\caption{a) FUGIN $^{12}$CO(J =1$-$0) map of intensity (moment-0) in the direction of the selected area around {\it l} = 13$\degr$.7 -- 14$\degr$.9 and {\it b} = $-$0$\degr$.5 -- +0$\degr$.1.
b) FUGIN $^{13}$CO(J =1$-$0) map of intensity (moment-0). A shell-like feature is also highlighted in the panel (see a broken ellipse in white). 
c) FUGIN C$^{18}$O(J =1$-$0) map of intensity (moment-0). Some selected molecular clumps are shown by 
squares. 
In each moment-0 map, the molecular emission is integrated from 31.6 to 46 km s$^{-1}$.}
\label{fig9}
\end{figure*}

\begin{figure*}
\epsscale{0.52}
\plotone{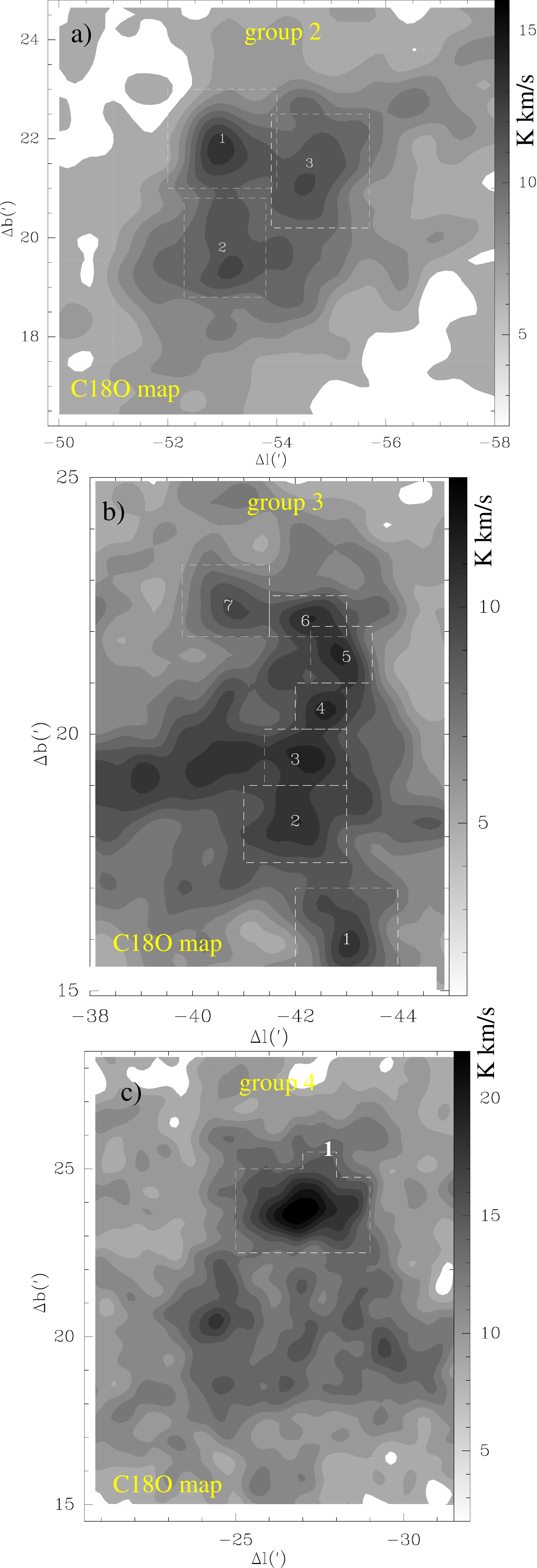}
\caption{Selected molecular clumps in the direction of group2, group3, and 
group4 using the FUGIN C$^{18}$O(J =1$-$0) map (see Figure~\ref{fig9}c and also Table~\ref{moltab}). 
In each panel, the axes are offsets (in arcmin) with respect to the central position (i.e., {\it l} = 14$\degr$.901; 
{\it b} = $-$0$\degr$.499).} 
\label{mfg}
\end{figure*}

\begin{figure*}
\epsscale{0.87}
\plotone{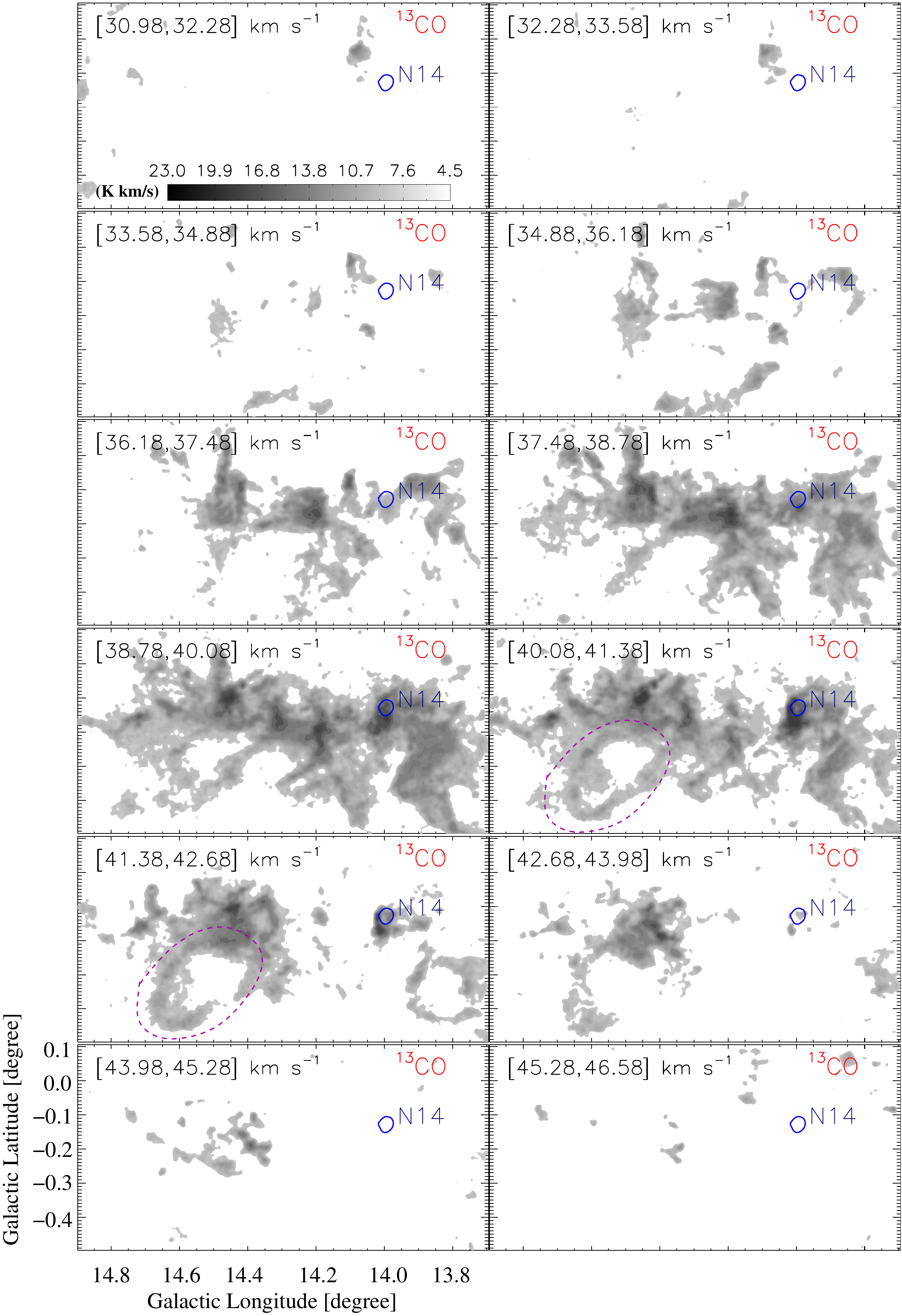}
\caption{The integrated velocity channel maps of $^{13}$CO(J =1$-$0) (at velocity intervals of 1.3 km s$^{-1}$). 
The bubble N14 is indicated by the MAGPIS 20 cm continuum contour (in blue) with a level of 3.7 mJy beam$^{-1}$. 
The $^{13}$CO contours are shown with the levels of 4.5, 5, 6, 7, 8, 9, 10, 11, 
13, 15, 18, 21, and 23 K km s$^{-1}$. A shell-like feature is also highlighted in two panels (see a broken ellipse in magenta). 
The gray-scale bar in the first left panel is applicable to all the other maps.}
\label{fig10}
\end{figure*}

To study the distribution of molecular gas, the integrated FUGIN $^{12}$CO, $^{13}$CO, and C$^{18}$O intensity 
maps are presented in Figures~\ref{fig9}a,~\ref{fig9}b, and~\ref{fig9}c, respectively. 
In each intensity map, the molecular gas is integrated over a velocity range of [31.6, 46] km s$^{-1}$ (see also Figure~\ref{zfig1}). 
In the direction of the extended physical system, the distribution of 
molecular gas allows us to infer the existence of a GMC (extension $\sim$59 pc $\times$ 29 pc), which contains several dense regions traced using the C$^{18}$O gas. 
In Figure~\ref{fig9}b, a shell-like feature is highlighted by a broken ellipse, and is prominently seen in the $^{13}$CO map. 
Using the C$^{18}$O line data, we have selected 11 molecular clumps in the direction of the 
bubble N14 (see three clumps in ATLASGAL group2), G014.194$-$00.194 (see seven clumps in ATLASGAL group3), 
and  G14.427$-$00.075 (see one clump in ATLASGAL group4) (see squares in Figure~\ref{fig9}c). 
Using the zoomed-in maps of C$^{18}$O, Figures~\ref{mfg}a,~\ref{mfg}b, and~\ref{mfg}c show the position(s) of the selected molecular clump(s) in the 
direction of group2, group3, and group4, respectively (see also Table~\ref{moltab}). 
In Figure~\ref{fig10}, we display the integrated $^{13}$CO 
velocity channel maps (at intervals of 1 km s$^{-1}$), 
revealing several clumpy regions in the GMC. In each velocity panel, the location of the bubble N14 is highlighted 
by a radio continuum contour. The shell-like feature is also marked in two velocity channel panels by a broken ellipse. 

Using the optically thin C$^{18}$O line data, the total molecular masses ($M_\mathrm{mc}({\mathrm H_2})$) 
and the virial masses ($M_\mathrm{vir}$) for the molecular clumps highlighted 
in Figures~\ref{fig9}c and~\ref{mfg} are estimated. 
For the calculations, we have used the procedures and equations given in \citet{dewangan19} 
\citep[see also][for equations]{mangum16,frerking82,maclaren88}. 
Adopting the values of mass and clump diameter for each molecular clump, the mean 
number density ($\bar n$) is also estimated. 
The derived physical parameters are tabulated in Table~\ref{moltab}, which shows that 
all of these molecular clumps are massive ($>$10$^{3}$ M$_{\odot}$) and dense ($>$10$^{4}$ cm$^{-3}$).
One can notice that the value of $M_\mathrm{vir}$ is calculated for the case of 
a spherically symmetric clump with a constant density, no external pressure, and no magnetic field. 
Our calculations enable us to determine the ratio of $M_\mathrm{mc}$ and $M_\mathrm{vir}$ for 
all the selected molecular clumps (see Table~\ref{moltab}). 
Note that the uncertainties of both mass estimates are the combinations of several factors \citep[e.g.,][]{dewangan19}, some of which 
are unknown (such as, clump density profiles, the C$^{18}$O excitation temperature etc.). 
We can consider an uncertainty in the mass calculation to be typically $\sim$20\% and at largest $\sim$50\%. 
Taking into account a value of $\sim$50\% for both mass uncertainties, one could conclude that at least five of the 
eleven C$^{18}$O clumps (two in the group2, two in the group3 and one in the group4) 
with $M_\mathrm{mc}$/$M_\mathrm{vir}$ $\geq$ 2 should be unstable. 
It implies that these molecular clumps are unstable against gravitational collapse.

\begin{figure*} 
\epsscale{1.2}
\plotone{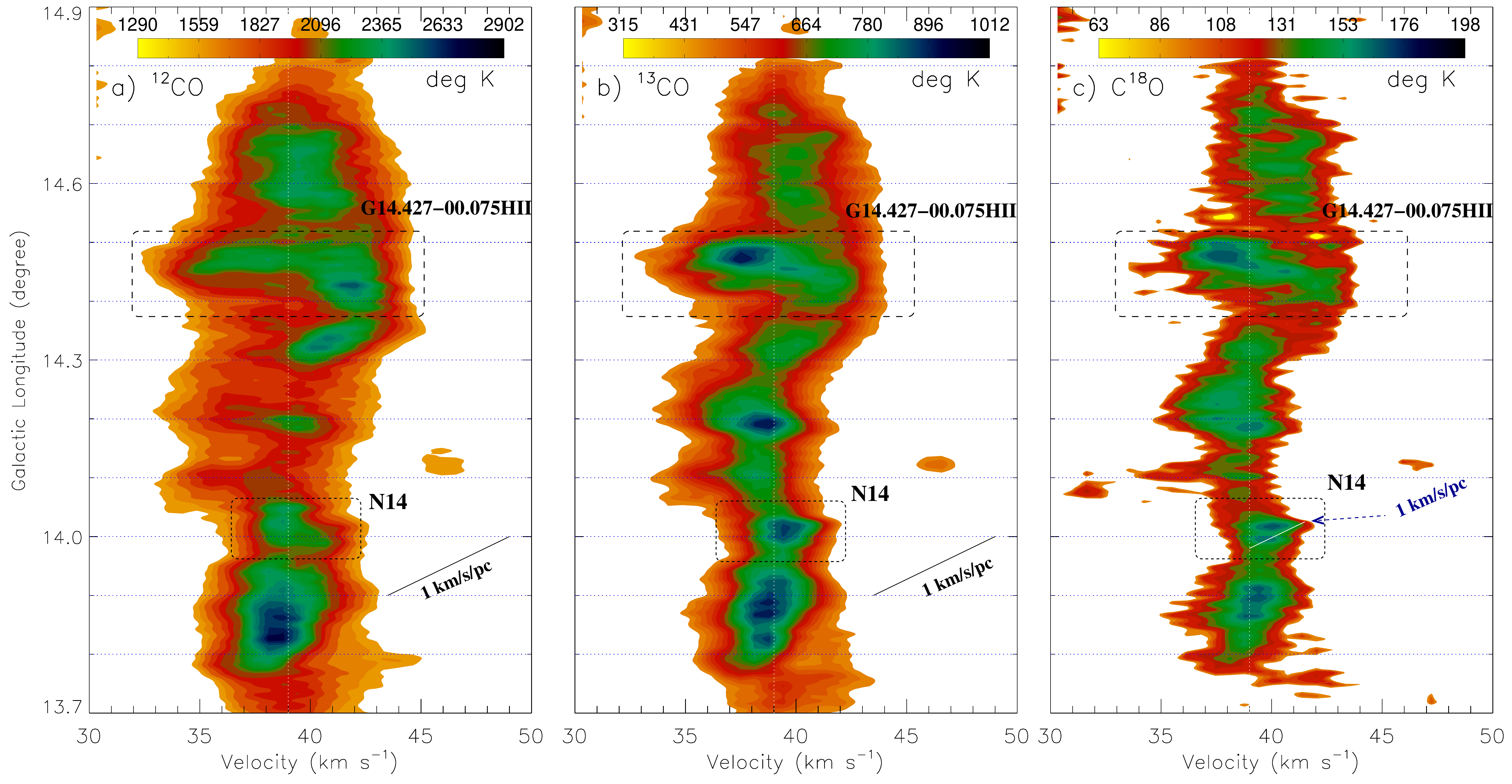}
\epsscale{1.2}
\plotone{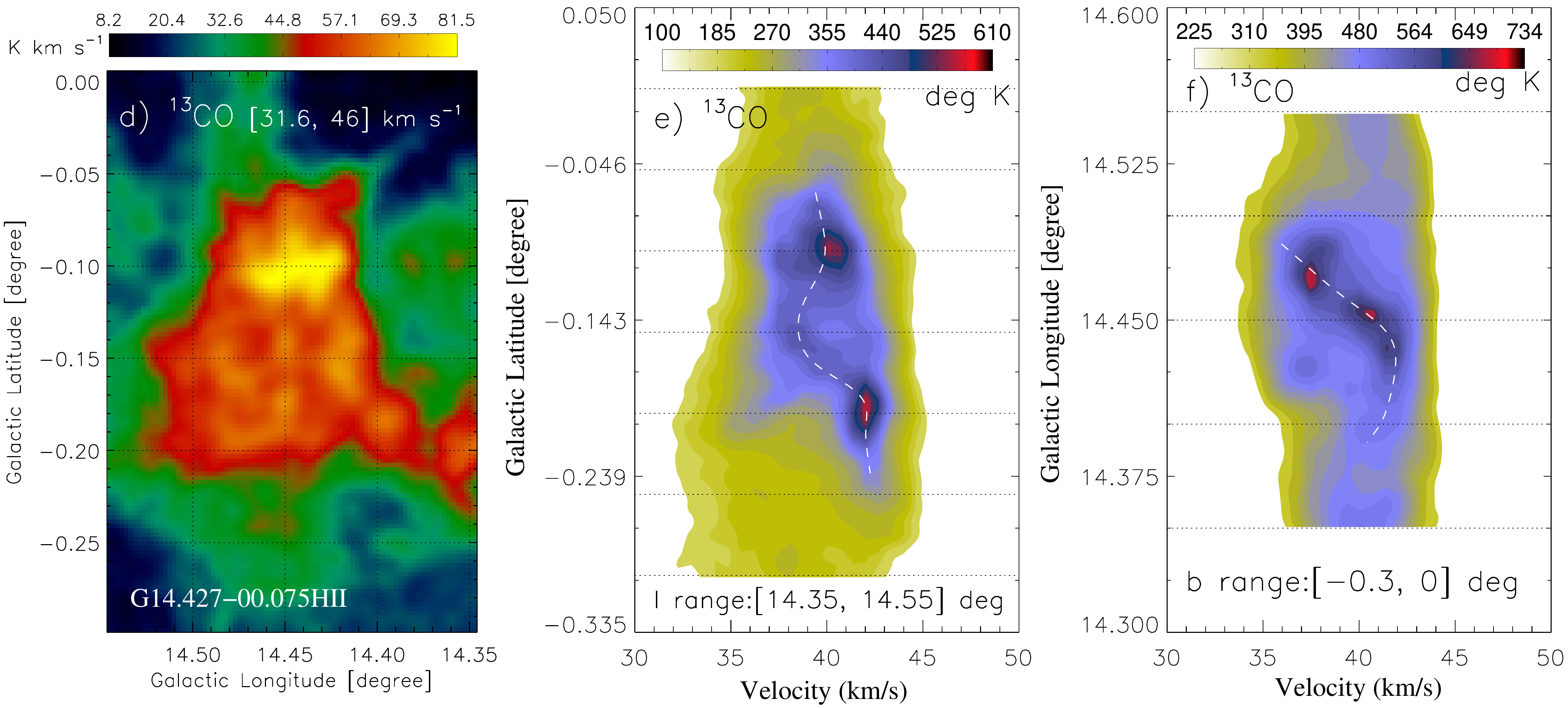}
\caption{Longitude-velocity maps of a) $^{12}$CO, b) $^{13}$CO and c) C$^{18}$O.  
d) FUGIN $^{13}$CO integrated intensity map toward G14.427-00.075HII. e) Latitude-velocity map of $^{13}$CO in the direction of G14.427-00.075HII. The molecular emission is integrated over the longitude range, which is highlighted in the panel. 
f) Longitude-velocity map of $^{13}$CO toward G14.427-00.075HII. The molecular emission is integrated over the latitude range, which is labeled in the panel. 
In panels a), b) and c), the molecular emission is integrated over the latitude range from $-$0$\degr$.5 to 0$\degr$.1, and 
a scale bar corresponding to 1 km s$^{-1}$ pc$^{-1}$. 
In panels a), b) and c), the observed velocity structures toward the bubble N14 and G14.427-00.075HII are highlighted by a dotted box and a dashed box, respectively. A scale bar corresponding to 1 km s$^{-1}$ and 1 pc in the horizontal and vertical axes, respectively, is shown in panels a) and b), while a slope of 1 pc per km s$^{-1}$ is in panel c).} 
\label{fig11}
\end{figure*}

Figures~\ref{fig11}a,~\ref{fig11}b, and~\ref{fig11}c display the longitude-velocity maps of $^{12}$CO, $^{13}$CO, 
and C$^{18}$O, respectively. These molecular emissions exhibit a large spread in velocities over a range of 
$\sim$10--12 km s$^{-1}$ over the entire physical system. 
In the direction of the extended physical system, continuous velocity structures are seen, where velocity gradients are also evident. 
The velocity appears to be oscillating along the longitude in all the position-velocity maps. 
Overall, the analysis of molecular gas confirms the spatial and velocity connections of all the selected groups. 
It indicates that due to some physical processes, the extended physical system breaks into smaller groups in 
a hierarchical manner (see Section~\ref{sec:disc} for more details). 

In the integrated molecular map of $^{13}$CO, we have highlighted a shell-like feature toward {\it l} = 14$\degr$.3--14$\degr$.5 or the site G14.427$-$00.075HII. 
In this selected longitude direction or the site G14.427$-$00.075HII, an arc-like feature is seen in velocity (see a dashed box in 
Figures~\ref{fig11}a,~\ref{fig11}b, and~\ref{fig11}c). In Figure~\ref{fig11}d, we display a zoomed-in view of the cloud associated with the site G14.427$-$00.075HII. 
To further examine the velocity structure toward the site G14.427$-$00.075, Figures~\ref{fig11}e and~\ref{fig11}f present the latitude-velocity and longitude-velocity 
maps of $^{13}$CO, respectively. The latitude-velocity map clearly reveals the arc-like feature in the velocity space, which is also seen in Figure~\ref{fig11}f 
(see a broken curve in Figures~\ref{fig11}e and~\ref{fig11}f). 
Hence, both the position-velocity maps favour the presence of an expanding H\,{\sc ii} region toward the site G14.427$-$00.075.  
Earlier, the semi-ring-like or C-like or arc-like structure in velocity has been observed in massive star-forming 
regions (such as, Orion nebula \citep{wilson05}, Perseus molecular cloud \citep{arce11}, W42 \citep{dewangan15}, 
S235 \citep{dewangan16}). Using a modeling of expanding bubbles in a turbulent medium, 
\citet{arce11} proposed that an expanding shell associated with the cloud should be responsible for the 
semi-ring-like or C-like structure in velocity. 
Considering the signature of the expanding H\,{\sc ii} region, the observed radio continuum emission as well as the 
extended temperature structures, we suggest the impact of massive star(s) associated with the site G14.427$-$00.075HII to its vicinity.

The noticeable velocity gradient (i.e., $\sim$1 km s$^{-1}$ pc$^{-1}$) is also clearly seen toward 
the bubble N14 (see a solid white line in Figure~\ref{fig11}c), where the extended and spherical-like temperature feature 
is evident (see Figure~\ref{fig6}a). 
Earlier, \citet{yan16} reported the molecular maps toward the bubble N14 using different molecular lines (see Figure~5 in their paper). 
\citet{sherman12} published the observational data at 3.3 mm continuum and several 
molecular lines toward the bubble N14. Based on the N$_{2}$H$^{+}$ line data, 
they pointed out that the bubble N14 is expanding into a very inhomogeneous cloud. Our findings favour this interpretation.  

\begin{figure*} 
\epsscale{1}
\plotone{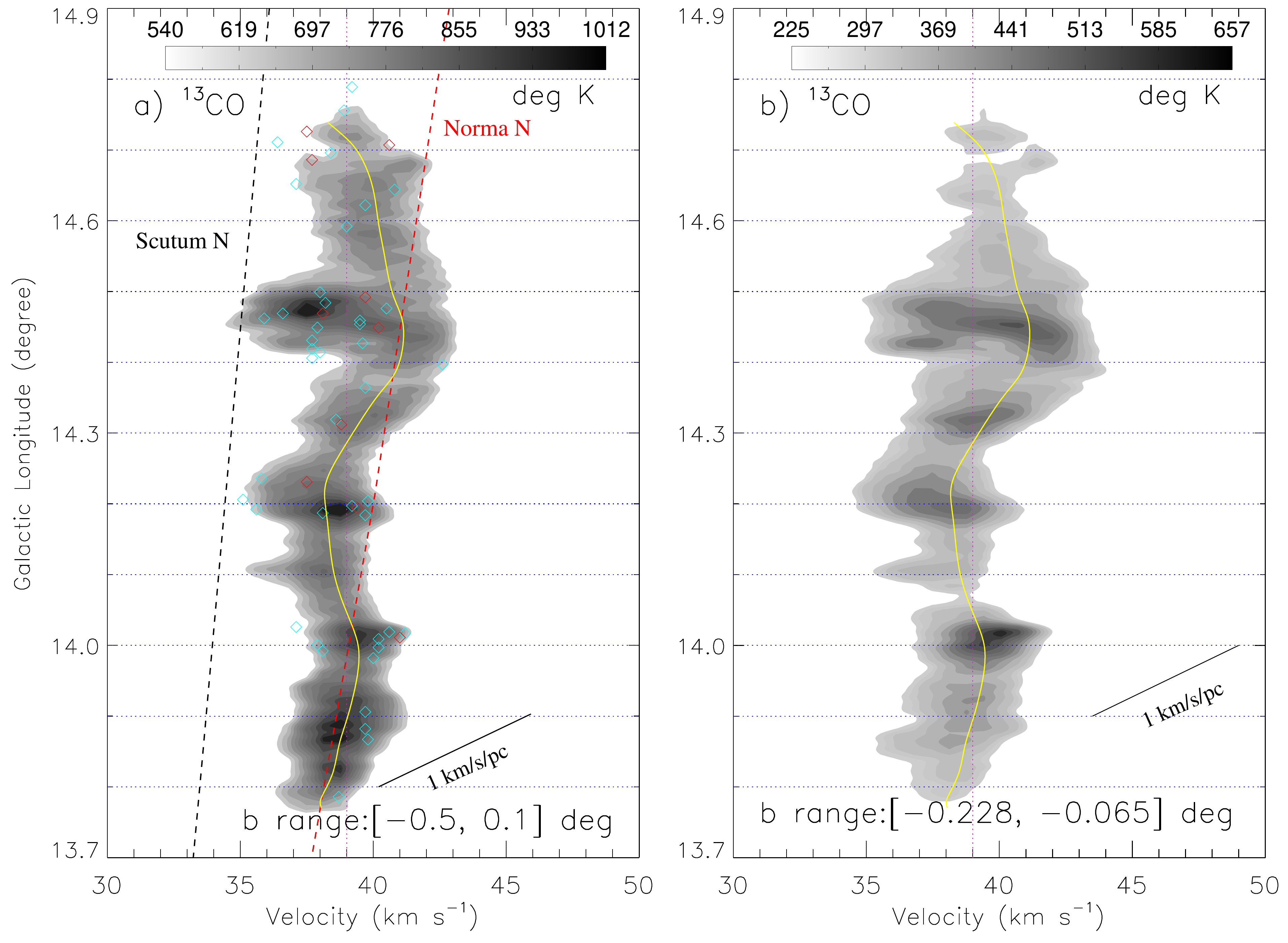}
\epsscale{1}
\plotone{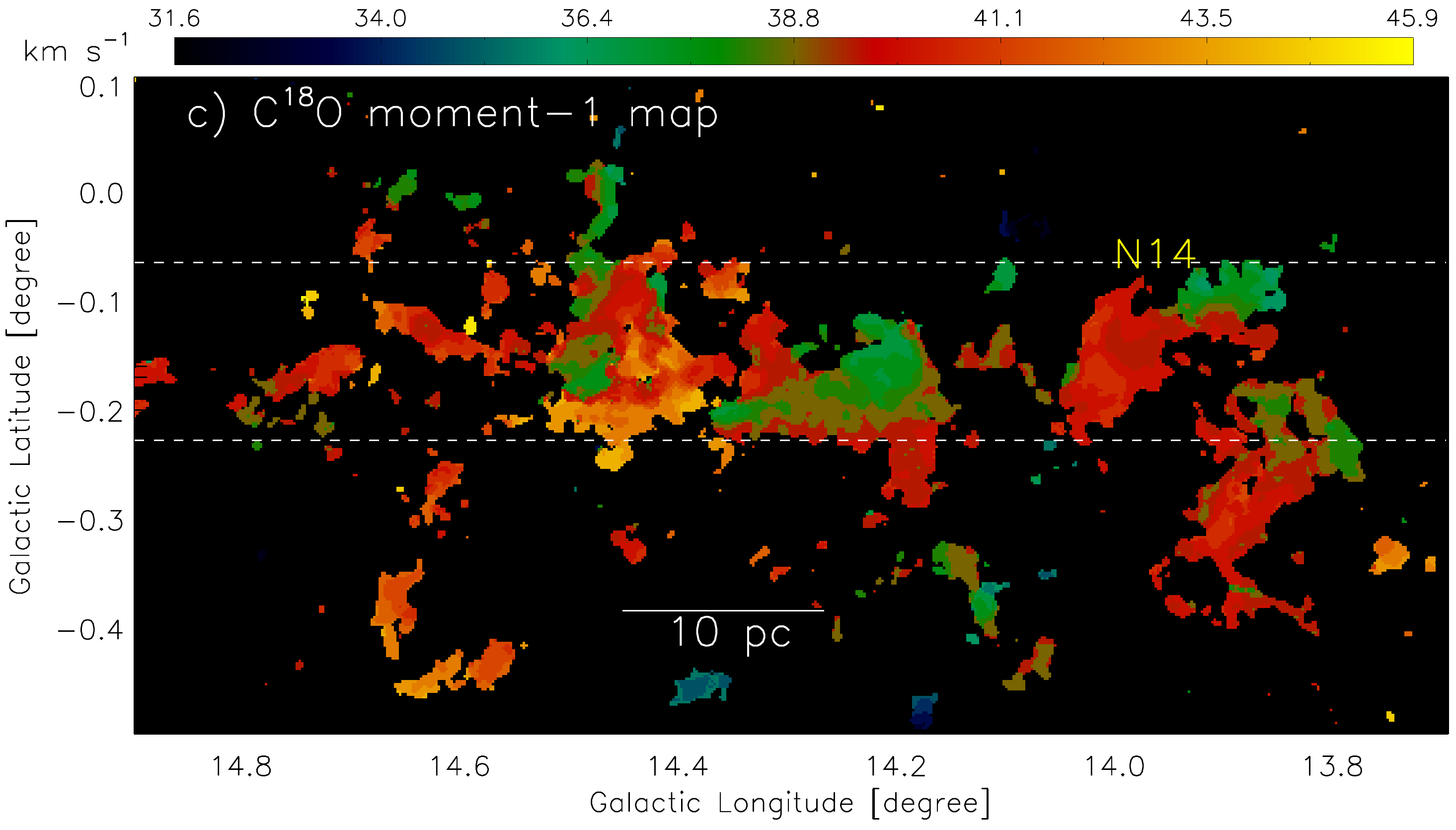}
\caption{a) Longitude-velocity map of $^{13}$CO (see also Figure~\ref{fig11}b).  
The $^{13}$CO emission is integrated over the latitude from $-$0$\degr$.5 to 0$\degr$.1, and is shown between 540~and~1012~deg~K. 
The panel also shows the Scutum and Norma arms \citep[from][]{reid16}. 
The near side of the arms is presented by broken curves. 
The $V_\mathrm{lsr}$ of each ATLASGAL clump against its longitude is also displayed in the 
plot (see diamonds and also Figure~\ref{fig1}a). 
\citet{urquhart18} reported the NH$_{3}$ line-width for some ATLASGAL clumps, which are 
highlighted by red diamonds in the figure (see also Table~\ref{tab3}). 
b) Longitude-velocity map of $^{13}$CO. The $^{13}$CO emission is integrated over the latitude from $-$0$\degr$.228 to $-$0$\degr$.065. 
In panels a) and b), an arbitrarily chosen solid curve (in yellow) shows an oscillatory-like velocity pattern along 
the longitude, and a scale bar corresponding to 1 km s$^{-1}$ pc$^{-1}$ is also shown. 
c) Intensity-weighted mean velocity map (or the first moment map) of C$^{18}$O. The color bar indicates the mean velocity (in km s$^{-1}$). 
Broken lines are shown at {\it b} = $-$0$\degr$.228 and $-$0$\degr$.065.}
\label{fig12}
\end{figure*}

In order to highlight the oscillatory-like velocity pattern, in Figures~\ref{fig12}a and~\ref{fig12}b, an arbitrarily 
chosen curve is drawn in the longitude-velocity maps of $^{13}$CO. Figure~\ref{fig12}a is the same as in Figure~\ref{fig11}b, 
but the $^{13}$CO emission is shown for higher contour levels. In Figure~\ref{fig12}b, the molecular emission is 
integrated over a small range of latitude (i.e., $-$0$\degr$.228 to $-$0$\degr$.065). 
In this selected latitude range, most of the molecular emission is observed toward the extended physical system (see broken lines in Figure~\ref{fig12}c). 
In addition to the oscillatory-like velocity pattern, in Figure~\ref{fig12}b, velocity gradients are 
also evident toward the selected groups/sub-regions, as discussed above.
In Figure~\ref{fig12}c, we display the first moment map of C$^{18}$O, showing the intensity-weighted mean velocity of the emitting gas. 
In the first moment map, one can clearly find noticeable velocity spread in the direction of the selected groups/sub-regions. 
We have also shown the distribution of the $V_\mathrm{lsr}$ of the ATLASGAL 
clumps and the locations of Galactic arms in Figure~\ref{fig12}a. 
We also find the information of the NH$_{3}$ line-widths toward at least 10 ATLASGAL 
clumps (see red diamonds in Figure~\ref{fig12}a and also Table~\ref{tab3}) from \citet{urquhart18}, 
which are used to compute the virial masses of the clumps. 
Based on the ratio of $M_\mathrm{clump}$ and $M_\mathrm{vir}$, we find that 
these clumps (with $M_\mathrm{clump}$ $>$ $M_\mathrm{vir}$) are unstable 
against gravitational collapse. 
\subsection{Temperature map, column density map, and embedded protostars}
\label{sec:hermap} 
\begin{figure*}
\epsscale{1}
\plotone{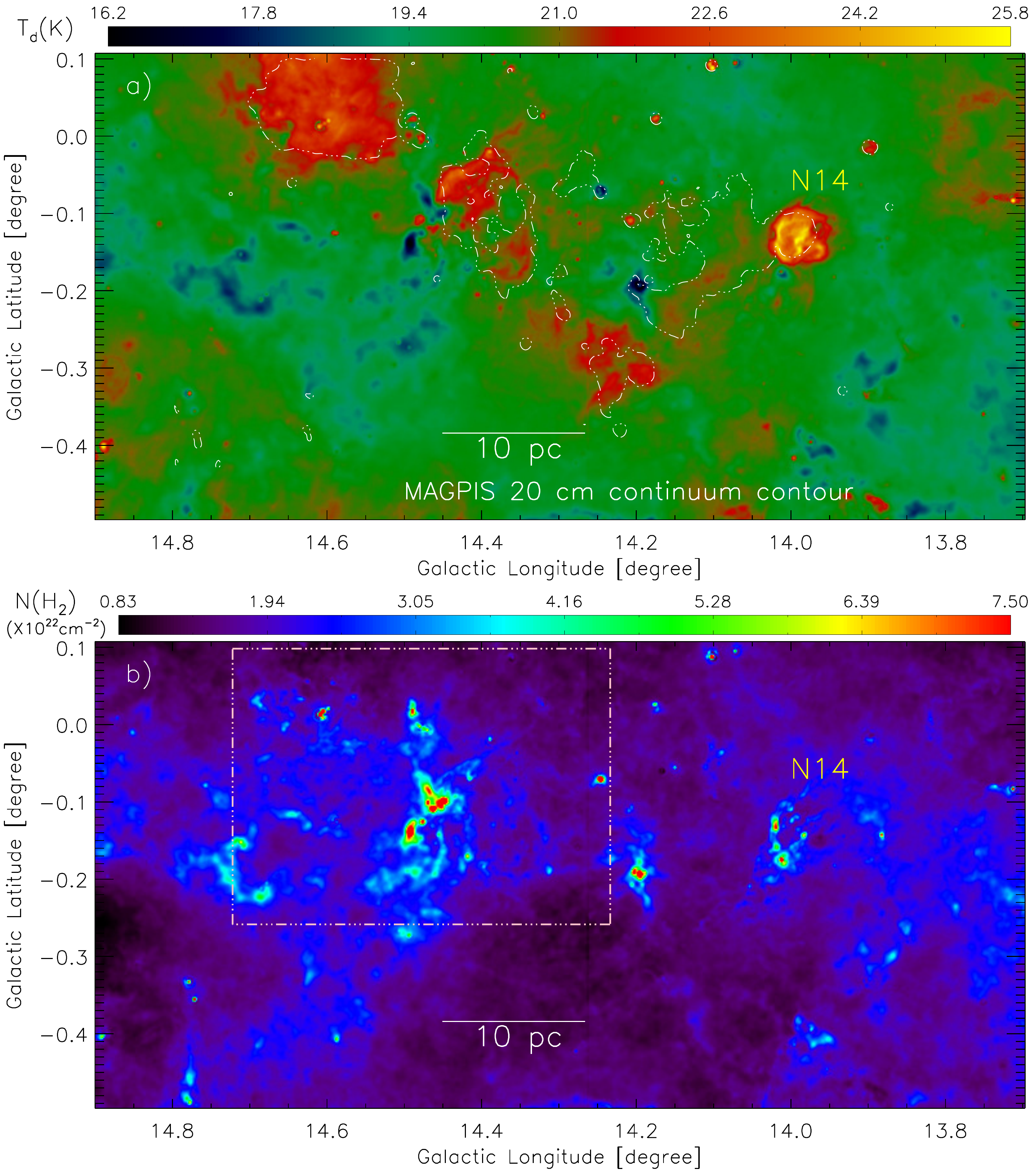}
\caption{a) {\it Herschel} temperature map of the selected area toward {\it l} = 13$\degr$.7 -- 14$\degr$.9 and {\it b} = $-$0$\degr$.5 -- +0$\degr$.1. 
The MAGPIS contour with a level of 2.2 mJy beam$^{-1}$ is also shown in the figure. 
b) {\it Herschel} column density ($N(\mathrm H_2)$) map of the selected area in this paper. 
The dotted-dashed box encompasses the area shown in Figures~\ref{fig6a}a,~\ref{fig6a}b, 
and~\ref{fig6a}c.}
\label{fig6}
\end{figure*}

In Figures~\ref{fig6}a and~\ref{fig6}b, we have presented the {\it Herschel} temperature and column 
density ($N(\mathrm H_2)$) maps (resolution $\sim$12$''$) of our selected target area, respectively. 
These maps\footnote[2]{http://www.astro.cardiff.ac.uk/research/ViaLactea/} were generated for the {\it EU-funded ViaLactea project} \citep{molinari10b} 
using the Bayesian {\it PPMAP} method \citep{marsh15,marsh17}, which was applied on the {\it Herschel} 
images at wavelengths of 70, 160, 250, 350 and 500 $\mu$m. In Figure~\ref{fig6}a, the {\it Herschel} temperature 
map is also superimposed with the MAGPIS 20 cm continuum contour. 
Radio continuum emission or H\,{\sc ii} regions are found toward the areas with a relatively warm dust emission ($T_\mathrm{d}$ $>$21~K). 
In the extended physical system, at least three highlighted sites (i.e., G14.71$-$0.19, G14.47$-$0.20, 
and G014.194$-$00.194) are associated with the areas of cold dust 
emission (i.e., $T_\mathrm{d}$ $\sim$16.5--18~K; see also Figure~\ref{fig1}a). 
The most prominent feature in the {\it Herschel} temperature map is seen toward the bubble N14.  
The temperature structure of the bubble N14 is almost spherical, which is in agreement with the radio morphology. 
It seems that the feedback from massive stars (such as, stellar wind, ionized emission, and radiation pressure) might 
have heated the surroundings and is responsible for the extended temperature structure. 

\begin{figure*}
\epsscale{0.75}
\plotone{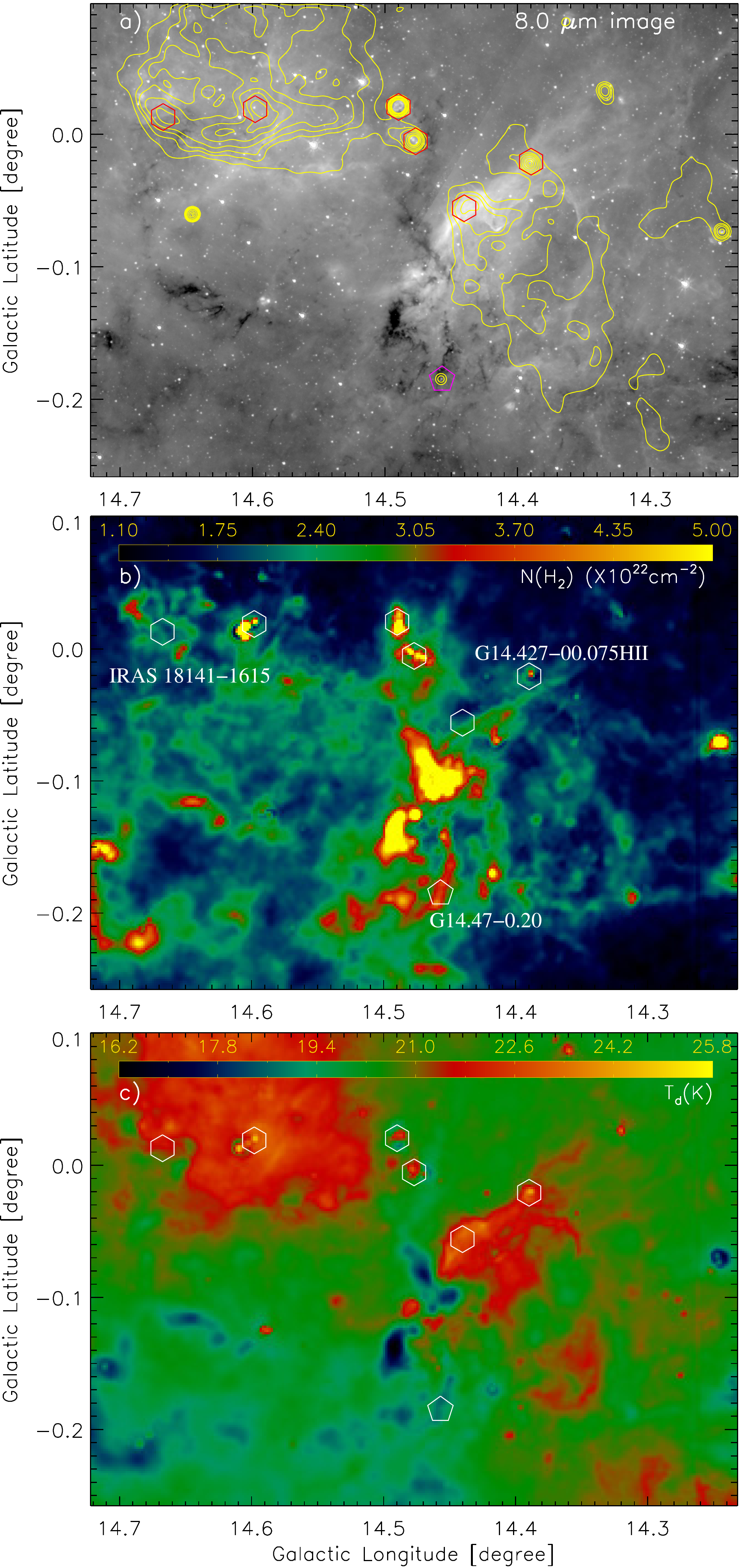}
\caption{A zoomed-in view of the area highlighted by a broken box in Figure~\ref{fig6}b.
a) Overlay of the MAGPIS contours and the THOR radio sources 
on the {\it Spitzer} 8.0 $\mu$m image. 
The MAGPIS contours are shown with the levels of 2.2, 2.8, 3.3, 4.0, 5.0, 6.0, and 6.7 mJy beam$^{-1}$. 
b) Overlay of the THOR radio sources on the {\it Herschel} column density 
map.
c) Overlay of the THOR radio sources on the {\it Herschel} temperature map. In each panel, hexagon and pentagon symbols are 
the same as in Figure~\ref{fig1}b.}
\label{fig6a}
\end{figure*}

The column density map shows the distribution of materials with high column densities ($>$ 2.4 $\times$ 10$^{22}$ cm$^{-2}$) toward 
the highlighted sites (e.g., N14, G014.194$-$00.194, 
G14.427$-$00.075HII (and G14.47$-$0.20), G14.71$-$0.19) in the extended physical system. 
Using the {\it Spitzer} 8.0 $\mu$m image, {\it Herschel} column density map, and 
{\it Herschel} temperature map, Figures~\ref{fig6a}a,~\ref{fig6a}b, and~\ref{fig6a}c display a zoomed-in view of 
the area hosting some highlighted sources (e.g., IRAS 18141$-$1615, G14.427$-$00.075HII, and G14.47$-$0.20), respectively. 
The positions of the THOR radio sources are marked in 
Figures~\ref{fig6a}a,~\ref{fig6a}b, and~\ref{fig6a}c. 
The MAGPIS 20 cm continuum contours are also overplotted on the 
{\it Spitzer} 8.0 $\mu$m image. The {\it Spitzer} image reveals extended bright emission as well as 
infrared dark clouds (IRDCs). The IRDCs are depicted as the absorption features against 
the Galactic background in the 8.0 $\mu$m image. The IRDCs are found with cold dust emission as well as 
high column density materials (see Figures~\ref{fig6a}a,~\ref{fig6a}b, and~\ref{fig6a}c). 
Previously, the extended emission at {\it Spitzer} 8.0 $\mu$m has been observed toward the bubble N14 \citep{churchwell06,dewangan13,yan16}. 
In general, it is known that the {\it Spitzer} 8.0 $\mu$m band contains polycyclic aromatic hydrocarbon (PAH) features at 7.7 and 8.6 $\mu$m. 
Considering the extended warm dust emission, PAH features and molecular gas surrounding the ionized emission, one can infer 
the existence of photon dominant regions (PDRs) in the extended physical system. The PDRs are traced by the PAH emission, 
and indicate the molecular/ionized gas interface where one can expect the influence of UV radiation liberated by a nearby massive star.

\begin{figure*}
\epsscale{0.92}
\plotone{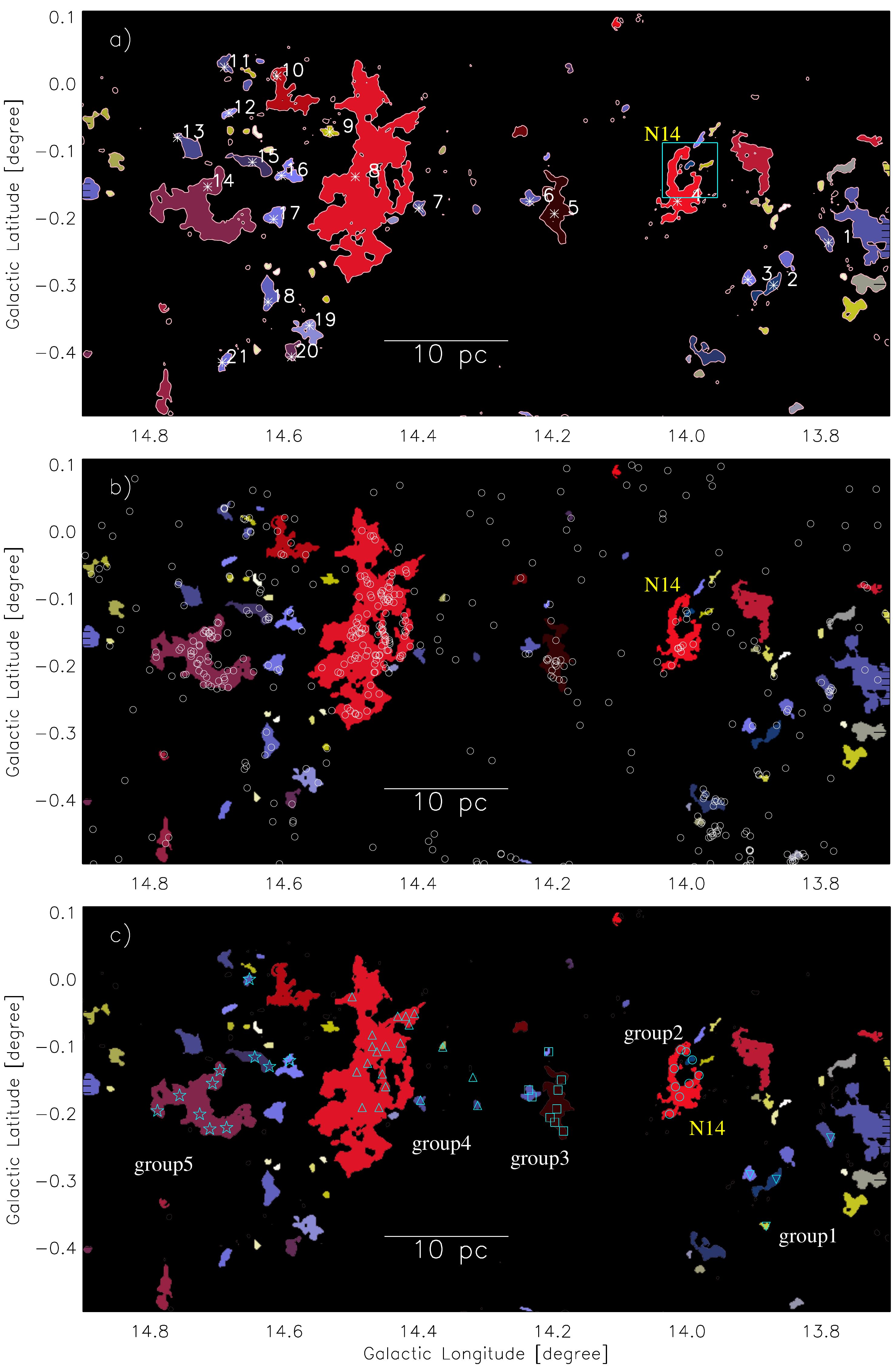}
\caption{a) The boundaries of several extended regions are shown in the {\it Herschel} column density map using the $N(\mathrm H_2)$ contour level of 2.4 $\times$ 10$^{22}$ cm$^{-2}$. 
Each selected extended region is highlighted by an asterisk along with its corresponding clump ID (see Table~\ref{ftab1}). 
The cyan box area is zoomed-in Figures~\ref{fig4}a,~\ref{fig4}b,~\ref{fig4}c, and~\ref{fig4}d. 
b) The positions of the identified Class~I YSOs are shown toward the extended regions traced in the {\it Herschel} column density map. 
In this paper, the Class~I YSOs, highlighted by open circles, are selected using the {\it Spitzer} color-color plot ([4.5]$-$[5.8] vs [3.6]$-$[4.5]; see text for more details). 
c) The positions of the ATLASGAL clumps at 870 $\mu$m are displayed toward the extended regions traced in Figure~\ref{fig7}a. Different symbols are the same as in Figure~\ref{fig2}a.}
\label{fig7}
\end{figure*}

To infer different sub-regions in the extended physical system, 
the {\it clumpfind} algorithm is adopted with the $N(\mathrm H_2)$ 
of 2.4 $\times$ 10$^{22}$ cm$^{-2}$ as an input parameter. 
At least 21 sub-regions are identified, which are also marked and labeled 
in Figure~\ref{fig7}a. The total mass of each sub-region is estimated using 
an equation, $M_{area} = \mu_{\rm H_2} m_{\rm H} Area_{pix} \Sigma N(\mathrm H_2)$, where $\mu_{\rm H_2}$ (= 2.8) 
is defined earlier, $Area_{pix}$ is the area subtended by one pixel (i.e., 6$''$/pixel), 
and $\Sigma N(\mathrm H_2)$ is the total column density \citep[see also][]{dewangan17a}. 
In Table~\ref{ftab1}, the mass and the radius of each {\it Herschel} sub-region are listed. 
The masses of the {\it Herschel} sub-regions vary between 335 and 42975 M$_{\odot}$. 

Figure~\ref{fig7}b shows the distribution of the selected Class~I 
YSOs \citep[mean age $\sim$0.44 Myr;][]{evans09} toward the {\it Herschel} sub-regions, 
tracing the early phases of star formation activities in the extended physical system (see white circles).
In other words, star formation activities are traced toward all the groups of the ATLASGAL clumps. 
Previously, using the {\it Spitzer} 3.6--5.8 $\mu$m photometric data, \citet{hartmann05} and \citet{getman07} applied the infrared color conditions 
(i.e., [4.5]$-$[5.8] $\ge$ 0.7 mag and [3.6]$-$[4.5] $\ge$ 0.7 mag) to identify Class~I YSOs in a given star-forming region. 
We have used this selection scheme to select Class~I YSOs in the extended physical system. 
In this context, photometric magnitudes of point-like objects at 3.6--5.8 $\mu$m were obtained from the {\it Spitzer} 
GLIMPSE-I Spring' 07 highly reliable catalog. In this work, we considered only objects with a 
photometric error of less than 0.2 mag in the selected {\it Spitzer} bands. 

In Figure~\ref{fig7}c, one can examine the distribution of ATLASGAL clumps at 870 $\mu$m toward the {\it Herschel} 
sub-regions. A majority of ATLASGAL clumps at 870 $\mu$m (see Table~\ref{tab3}) are mainly found in the direction 
of four {\it Herschel} sub-regions (see IDs \# h4, h5, h8, and h14 in Table~\ref{ftab1}; mass 
range: 5400--42975 M$_{\odot}$), where signposts of active star formation are investigated (see Figure~\ref{fig7}c). 
Using the C$^{18}$O line data, dense molecular clumps (see Table~\ref{moltab}) are also identified toward 
the {\it Herschel} sub-regions (see IDs \# h4, h5, and h8 in Table~\ref{ftab1}). 
\subsection{GMRT radio continuum maps}
\label{subsec:radio}
\begin{figure*}
\epsscale{1}
\plotone{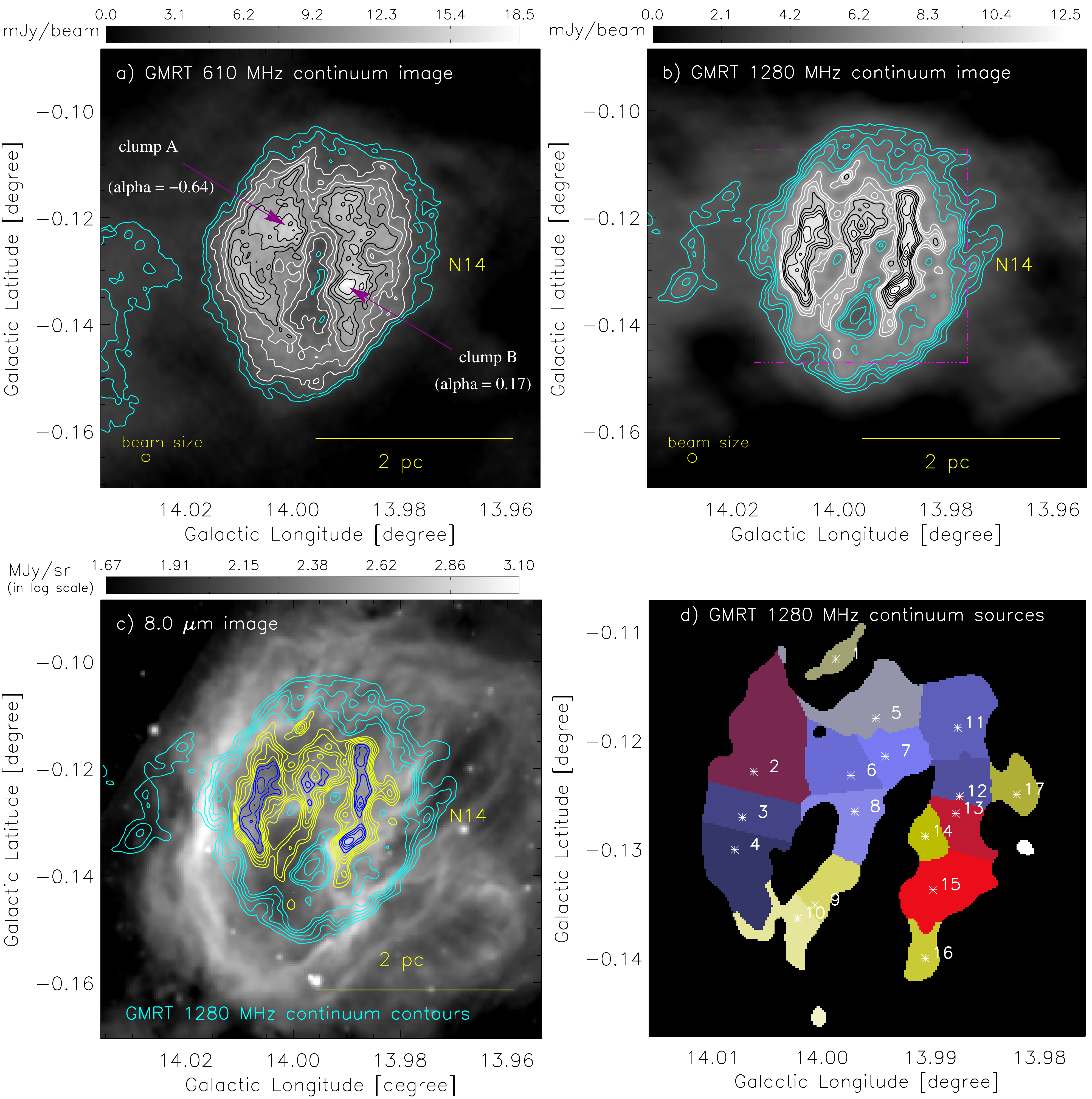}
\caption{a) A zoomed-in view of the bubble N14 using the GMRT 610 MHz 
continuum map (resolution $\sim$5$''$.56 $\times$ 5$''$.22). 
Cyan contours are 4 and 4.8 mJy beam$^{-1}$. White contours are 6.2, 8.0, and 9.5 mJy beam$^{-1}$. 
Black contours are 11, 12.5, 14, and 16 mJy beam$^{-1}$. b) A zoomed-in view of the bubble N14 using the GMRT 1280 MHz 
continuum map (resolution $\sim$6$''$). Cyan contours are 4.05, 4.5, 5.0, 5.5, and 6.0 mJy beam$^{-1}$. 
White contours are 7.6, 8.15, and 8.7 mJy beam$^{-1}$. 
Black contours are 9.25, 9.8, 10.35, 10.9, 11.45, 12.0, and 12.55 mJy beam$^{-1}$. The broken box (in magenta) encompasses the area shown 
in Figure~\ref{fig4}d. 
c) Overlay of the GMRT 1280 MHz continuum contours on 
the {\it Spitzer} 8.0 $\mu$m image. 
Cyan contours are 4.05, 4.5, 5.0, 5.5, and 6.0 mJy beam$^{-1}$. 
Yellow contours are 7.6, 8.15, 8.7, 9.25, and 9.8 mJy beam$^{-1}$.
Blue contours are 10.35, 10.9, 11.45, 12.0, and 12.55 mJy beam$^{-1}$.
d) The panel shows the boundaries of several radio continuum sources at 1280 MHz, 
which are identified using the radio continuum contours at 1280 MHz (see Figure~\ref{fig4}b). 
Each ionized clump is highlighted by an asterisk along with its corresponding clump ID. 
In panel a), two radio clumps ``A" and ``B" are identified using the GMRT radio continuum maps at 610 MHz and 1280 MHz.}
\label{fig4}
\end{figure*}

We find that the MAGPIS 20 cm continuum map reveals the compact radio continuum emission toward the bubble N14, while 
the diffuse radio emission is seen away from the bubble 
(see white contour in Figure~\ref{fig6}a). To further explore the ionized emission, low-frequency 
radio continuum maps of the bubble N14 are examined in this paper. 
In Figures~\ref{fig4}a and~\ref{fig4}b, we 
display high-resolution GMRT radio continuum maps at 610 MHz (beam size $\sim$5$''$.56 $\times$ 5$''$.22) 
and 1280 MHz (beam size $\sim$6$''$), respectively. 
The GMRT 610 MHz continuum map is also superimposed with the 610 MHz continuum contour (see Figure~\ref{fig4}a). 
In Figure~\ref{fig4}b, the GMRT 1280 MHz continuum map is also overlaid with the 1280 MHz continuum contour. 
Figure~\ref{fig4}c displays the overlay of the GMRT 1280 MHz continuum emission contours on 
the {\it Spitzer} 8.0 $\mu$m image. The spherical-like radio morphology observed in the GMRT maps is found well within the bubble boundary. 
In Figures~\ref{fig4}a,~\ref{fig4}b, and~\ref{fig4}c, different color contours are used to show the inner 
radio morphology within the bubble N14. 

Using the GMRT 1280 MHz continuum map, at least 17 radio clumps are identified toward 
the bubble N14, which are shown in Figure~\ref{fig4}d (see a broken box in Figure~\ref{fig4}b). 
In this analysis, the {\it clumpfind} algorithm \citep{williams94} was employed. 
Adopting the similar procedures given in \citet{dewangan17a}, the Lyman continuum photons \citep[see also][for equation]{matsakis76} 
and spectral type of each radio source are computed. In the calculation, we used a distance of $\sim$3.1 kpc, an electron temperature of 10$^{4}$ K, and 
the models of \citet[][see his Table 2]{panagia73}. The analysis suggests that all the ionized clumps are 
powered by massive B-type stars. We have tabulated the derived physical properties of the ionized 
clumps (i.e., deconvolved effective radius of the ionized clump ($R_\mathrm{HII}$), total flux ($S{_\nu}$), 
Lyman continuum photons ($\log{N_\mathrm{uv}}$), dynamical age ($t_\mathrm{dyn}$), and radio spectral type) in Table~\ref{gtab2}. 
Using the equation given in \citet{dyson80}, we have computed the values of $t_\mathrm{dyn}$ for each 
radio clump \citep[see][for more details]{dewangan17a}. 
The ages of all these radio clumps are found between $\sim$10$^{3}$--10$^{4}$ yr 
for an initial particle number density (n$_{i}$) of 10$^{3}$ cm$^{-3}$, indicating that they are very young. 
Furthermore, the values of $S{_\nu}$ and $N_\mathrm{uv}$ are also computed for the entire extended spherical 
emission using the GMRT 1280 MHz continuum map. The integrated flux is 
estimated to be $S{_\nu}$ $\sim$3.54 Jy at 1280 MHz, which yields $\log{N_\mathrm{uv}}$ to be 
$\sim$ 48.42. It implies that the observed extended radio emission can be explained by a single ionizing star of spectral 
type O8.5V \citep[e.g.,][]{panagia73}. It is also in agreement with the analysis of 
the GMRT 610 MHz data (not presented here) as well as the previously reported value of 
$\log{N_\mathrm{uv}}$ \citep[= 48.36;][]{beaumont10}. 

\begin{figure*}
\epsscale{1}
\plotone{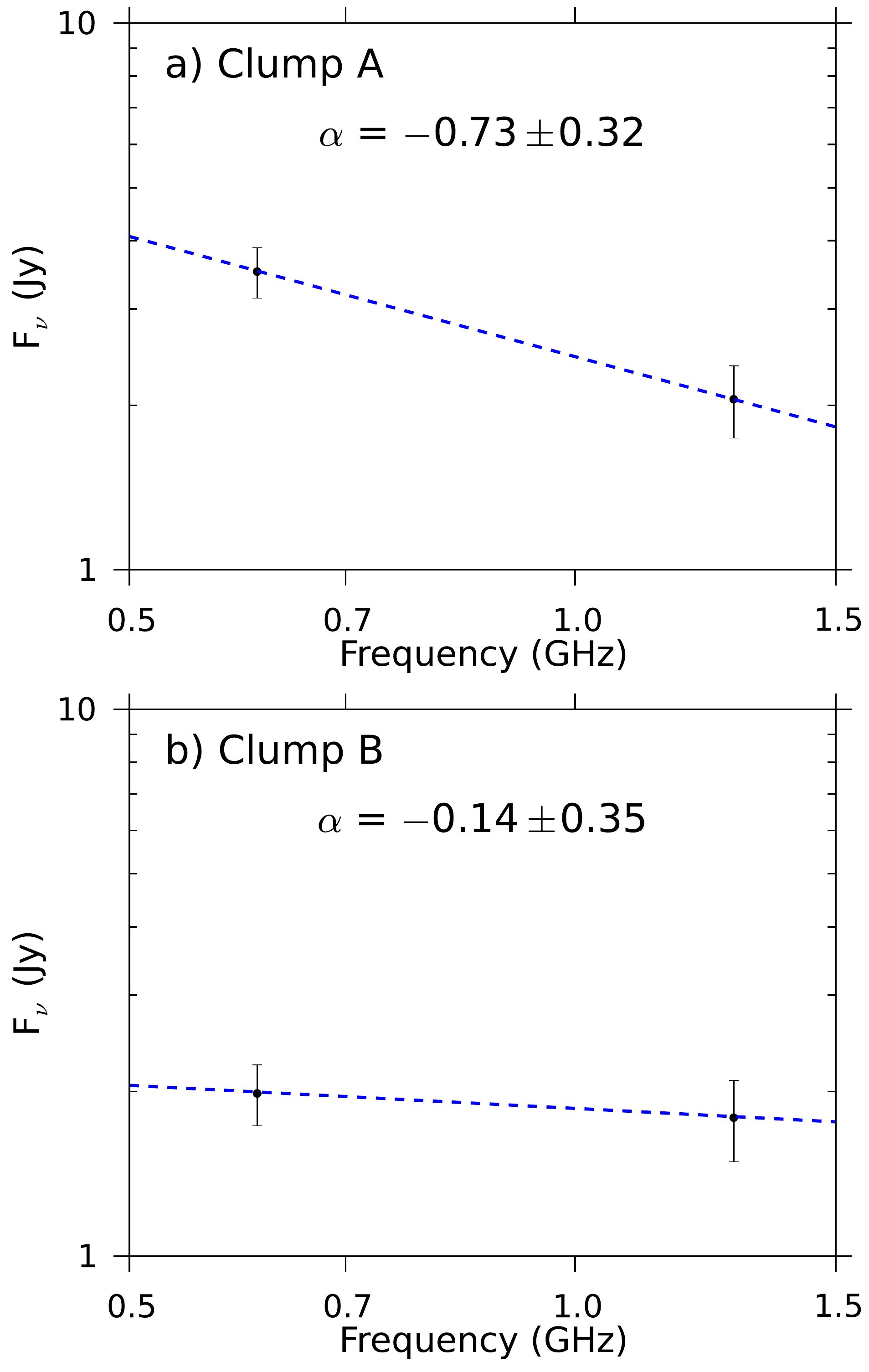}
\caption{a) Radio spectral index plot of the radio clump ``A", which is indicated in Figure~\ref{fig4}a.
b) Radio spectral index plot of the radio clump ``B" (see Figure~\ref{fig4}a). 
In each panel, filled circles (in black) are the flux densities at 610 and 1280 MHz.}
\label{fig5}
\end{figure*}

In general, radio continuum data at low frequencies are very useful to trace the non-thermal emission in a given astrophysical object \citep[e.g.,][]{becker18}. 
In order to infer the spectral indices of the radio clumps toward the bubble N14, the GMRT radio continuum 
maps at 610 and 1280 MHz are convolved to the same (lowest) resolution, and at least two radio 
clumps (i.e., clump A and clump B) are identified. 
Figures~\ref{fig5}a and~\ref{fig5}b show the radio spectral index plots of clump A and clump B, respectively. 
The spectral indices of the radio clumps are also labeled in Figures~\ref{fig5}a and~\ref{fig5}b. 
The positions of these clumps are also highlighted in Figure~\ref{fig4}a (see clump A ($\alpha$ $\sim$$-$0.73) and clump B ($\alpha$ $\sim$$-$0.14)).
This exercise indicates the presence of non-thermal emission in the bubble N14. 
Additionally, in the direction of {\it l} $>$ 13$\degr$.4, we also find that several H\,{\sc ii} regions or THOR radio sources show non-thermal emission.  
It has been reported that H\,{\sc ii} regions powered by massive OB stars are normally associated with thermal emission \citep[e.g.,][]{wood89,kurtz05,sanchez08,sanchez11,hoare12,purcell13,wang18,yang19}. 
In the literature, we also find some examples of H\,{\sc ii} regions (i.e., IRAS 17160$-$3707 \citep{nandakumar16} and IRAS 17256$-$3631 \citep{veena16}), 
which emit both thermal and non-thermal radiation.
The observed non-thermal emission in H\,{\sc ii} regions indicates the presence of relativistic electrons. 
More recently, it has been suggested that the non-thermal emission in H\,{\sc ii} regions might be referred to synchrotron radiation 
from locally accelerated electrons restrained in a magnetic field \citep[see][for more details]{padovani19}. 
\section{Discussion}
\label{sec:disc}
Based on the analysis of the molecular gas and the distribution of the ATLASGAL clumps, 
an extended physical system ($\sim$59 pc $\times$ 29 pc) is identified toward {\it l} = 13$\degr$.7--14$\degr$.9 
containing the bubble N14. The system is found to host at least five groups or sub-regions. 
The spatial and velocity connections of these sub-regions are also found in the analysis of the 
molecular line data (see Section~\ref{sec:coem}). 
Hence, the present work focuses to explore the physical processes responsible for the observed hierarchy in the extended physical system. 
These groups host unstable clumps (see Tables~\ref{tab3} and~\ref{moltab}), and are associated with the C$^{18}$O emission. 
These dense clumps are associated with Class~I YSOs, which trace the early phase of star formation (see Section~\ref{sec:hermap}). 
Furthermore, some of the 
sub-regions (e.g., group2 and group4) harbor massive OB stars, and their associated H\,{\sc ii} 
regions are found to be expanding in their surroundings (see Section~\ref{sec:coem}). 
In the direction of the H\,{\sc ii} regions, the extended structures in the {\it Herschel} temperature 
map are also found, illustrating the signatures of the impact of massive stars via their energetics (i.e., stellar wind, 
ionized emission, and radiation pressure). Hence, to explain the observed hierarchy, the 
application of ``globule squeezing" and ``collect and collapse" processes can be examined. 
These two processes are explained by the expanding H\,{\sc ii} regions powered by 
massive stars \citep[e.g.,][and references therein]{elmegreen77,whitworth94,elmegreen98,deharveng05,dale07,bisbas15,walch15,kim18,haid19}. 
 
In general, the impact of a massive star can be studied with the knowledge of the pressure of an H\,{\sc ii} region ($P_{HII}$), the 
radiation pressure ($P_{rad}$), and the stellar wind ram pressure ($P_{wind}$) \citep[e.g.,][]{dewangan15,dewangan17a}. 
Based on the values of different pressure components, the influence of massive stars to their surroundings is 
found to be more significant upto a projected distance of a few parsecs \citep[e.g.,][]{dewangan15,dewangan16b,dewangan17a,baug19}. 
Our observational results indicate that these processes might have influenced star formation activities locally in the extended physical system. 
However, these two processes may not explain 
the hierarchy extended upto about 59 pc in the selected target area. 

In one of the theoretical models, the observed star formation can be explained by convergent gas flows \citep{ballesteros99,heitsch08,vazqez07} or 
the CCC process \citep[e.g.,][]{habe92}. The model predicts that the convergence of streams of neutral gas can produce 
molecular clouds \citep[e.g.,][]{ballesteros99,heitsch08,vazqez07}. With time, one also expects the merging/converging/collision of the molecular clouds. 
However, we do not find distinct multiple velocity components associated with the extended physical system. In 
Figure~\ref{fig12}c, noticeable velocity spreads toward all the selected groups (except group5) are evident in the first moment map of C$^{18}$O. 
In Sections~\ref{subsec:radio1} and~\ref{sec:coem}, our observational results also show that the extended physical 
system is situated toward the near sides of the Scutum and Norma arms. 
Hence, locally, one cannot completely discard a triggered star formation scenario by convergent gas flows or 
the CCC process \citep[e.g.,][]{habe92}.

In recent years, it has been suggested that star-forming clouds seem to be in a 
state of global gravitational contraction \citep[e.g.,][]{hartmann12,VS+19}. It is also 
reported that the global collapse leads in a chaotic and hierarchical manner, yielding gravitationally 
driven fragmentation in star-forming molecular clouds \citep[e.g.,][]{BH04,HH08,Heitsch+09,Galvan+09,Peretto+13,Beuther+15,Liu+15,Liu+16b,Friesen+16,Jin+16,Hacar+17,Csengeri+17,
Yuan+18,Jackson+19,Barnes+19,VS+19}. The modeling results based on the global hierarchical gravitational collapse in molecular clouds by \citet{VS+19} indicate late birth 
of massive OB-stars and later loss of molecular gas feeding by feedback of massive stars. 
It also suggests that after the onset of global collapse, one can expect local collapse events in molecular clouds. 
Therefore, in the extended physical system, 
one can examine the observed hierarchy as the outcome of gravitational 
fragmentation \citep[see the review article by][]{hennebelle12}.  

In three-dimensional-models 
of molecular cloud formation in large-scale colliding flows including self-gravity, 
\citet{heitsch08} reported the formation of large-scale filaments due to global collapse of a molecular cloud. 
In the favour of this physical process, we find the important observational evidence in 
the velocity space of the molecular gas. 
The position-velocity maps of $^{12}$CO, $^{13}$CO, and C$^{18}$O show the oscillatory-like velocity pattern (with a period
of $\sim$8--13.5 pc and an amplitude of $\sim$2 km s$^{-1}$) in the direction of the entire longitude range (see Section~\ref{sec:coem}). The position-velocity maps have been produced for different ranges of latitude (i.e., {\it b} = [$-0\degr$.5, 0$\degr$.1], 
and  {\it b} = [$-$0$\degr$.228, $-$0$\degr$.065]; see Figures~\ref{fig12}a and~\ref{fig12}b). 
In the direction of the entire longitude range and {\it b} = [$-$0$\degr$.228, $-$0$\degr$.065], most of the molecular emission and IRDCs are observed (see broken lines 
in Figure~\ref{fig12}c). In the direction such selected area, the oscillatory-like velocity pattern is still evident (see Figure~\ref{fig12}b).
It implies that the extended physical system shows a sinusoidal-like (i.e. oscillatory) velocity structure with a significant fragmentation. 
In the direction of {\it l} = 14$\degr$.4--14$\degr$.5 or the G14.427$-$00.075HII region, an arc-like configuration is found in the velocity space, which 
shows signatures of an expansion (see Section~\ref{sec:coem}). 

Previously, in the case of the filament L1571 (length $\sim$0.35--0.70 pc at a distance of 144 pc), \citet{hacar11} performed a modeling of velocity oscillations as sinusoidal perturbations. Based on their analysis, the observed velocity oscillations along the filament (with a period of $\sim$0.19--0.24 pc and an amplitude of $\sim$0.04 km s$^{-1}$) were explained by the filament fragmentation process via accretion along the filament. L1517 is known as the site of low-mass star formation. Recently, \citet{dewangan19} also found an oscillatory pattern (with a period of $\sim$6--10 pc and an amplitude of $\sim$0.5 km s$^{-1}$) in velocity toward the S242 filamentary structure (length $\sim$30 pc at a distance of 2.1 kpc) hosting an H\,{\sc ii} region excited by a B-type star, and suggested the fragmentation of the filament. Figure~\ref{fig12}b in this paper looks similar to the published plot \citep[i.e., Figure~7b in][]{dewangan19} of S242, implying the onset of a similar fragmentation process. 
One can keep in mind that the extended physical system (at a distance of $\sim$3.1 kpc; size in longitude direction $\sim$59 pc) is not an extended filament, 
but it contains several IRDCs associated with the cold dust emission and high column density materials (see Section~\ref{sec:hermap}). 
It is also possible that the expansion of an H\,{\sc ii} region (i.e., local event) may diminish the collapse signature.
However, taking into account the existence of different groups/sub-regions in the extended physical system, the oscillatory pattern in velocity is suggestive of fragmentation process. 
It may also favour the multi-scale collapse, resulting a hierarchical configuration. 
We find the higher values of velocity amplitude and period of oscillations in the extended physical system compared to the filaments L1517 and S242, which could be explained by the existence of 
more massive and larger clumps associated with massive star formation at different stages of evolution. 
 
Taken together all these observed results, the concept of the global collapse seems to be applicable in the extended physical system, and 
can explain the presence of all spatially-distinct groups/sub-regions in the extended physical system.
\section{Summary and Conclusions}
\label{sec:conc}
The paper deals with a multi-wavelength study of a 
wide-scale environment toward {\it l} = 13$\degr$.7--14$\degr$.9 containing the bubble N14, allowing to examine the ongoing physical processes. 
The major outcomes of this work are presented below.\\

$\bullet$ The study of the FUGIN $^{12}$CO, $^{13}$CO, and C$^{18}$O gas at [31.6, 46] km s$^{-1}$ shows the presence of 
an extended physical system or molecular cloud (extension $\sim$59 pc $\times$ 29 pc) toward {\it l} = 13$\degr$.7--14$\degr$.9.\\
$\bullet$ 53 ATLASGAL 870 $\mu$m dust clumps at d $\sim$3.1 kpc are distributed toward the cloud. At least five spatially-distinct groups of the ATLASGAL clumps are 
selected through a visual inspection in the extended physical system.\\
$\bullet$ In the direction of four groups, gravitationally unstable clumps are identified, which are massive ($>$10$^{3}$ M$_{\odot}$) and dense ($>$10$^{4}$ cm$^{-3}$).\\
$\bullet$ Considering the distribution of Class~I YSOs (mean age $\sim$0.44 Myr), locally the 
early stage of star formation activity is observed toward each group of clumps.\\ 
$\bullet$ In the direction of the extended physical system, the position-velocity maps reveal continuous velocity structures, where velocity gradients are also evident.
The study of molecular gas displays the spatial and velocity connections of all the selected groups. 
These findings show a hierarchy in the extended physical system.\\
$\bullet$ The radio continuum and {\it Herschel} maps trace at least three groups of clumps 
associated with the expanding H\,{\sc ii} regions (including the bubble N14).\\
$\bullet$ In the direction of the H\,{\sc ii} regions, the warm dust emission ($T_\mathrm{d}$ $\sim$21--26 K) spatially 
coincides with the ionized emission. 
It implies that the molecular cloud appears to be influenced by massive OB stars.\\
$\bullet$ The observed spectral indices determined using the GMRT and THOR radio continuum data indicate the existence 
of non-thermal emission with the H\,{\sc ii} regions.\\
$\bullet$ The ionized emission traced in the GMRT 610 and 1280 MHz continuum maps shows an almost spherical 
morphology toward the bubble N14, which is found well within the bubble morphology depicted at 8.0 $\mu$m. 
A similar morphology is also observed in the {\it Herschel} temperature map (with $T_\mathrm{d}$ $\sim$21--26 K) toward the bubble N14. 
The ionizing photon flux values computed at both the GMRT bands refer to a single ionizing star of O8.5V spectral type.\\ 
$\bullet$ Using the GMRT 1280 MHz continuum map, at least 17 radio clumps, powered by B-type stars, are identified toward 
the bubble N14, which are found to be considerably young (age $\sim$10$^{3}$--10$^{4}$ years for n$_{i}$ = 10$^{3}$ cm$^{-3}$). \\\\
 
The analysis of molecular gas exhibits the 
oscillatory-like velocity pattern in the direction of the entire longitude range. Keeping in mind 
the presence of different groups/sub-regions, on a wide-scale, this velocity structure hints the 
onset of the fragmentation process in the extended physical system. 
Considering all the observational evidences presented in this work, the global collapse scenario seems to be operated in the extended physical system, which may also 
explain the observed hierarchy.
\acknowledgments  
We thank the anonymous reviewer for several useful comments and 
suggestions, which greatly improved the scientific contents of the paper.  
The research work at Physical Research Laboratory is funded by the Department of Space, Government of India. 
This work is based [in part] on observations made with the {\it Spitzer} Space Telescope, which is operated by the Jet Propulsion Laboratory, California Institute of Technology under a contract with NASA. 
This publication makes use of data from FUGIN, FOREST Unbiased Galactic plane Imaging survey with the Nobeyama 45-m telescope, a legacy project in the Nobeyama 45-m radio telescope. 
TB is supported by the National Key Research and Development Program of China through grant 2017YFA0402702. TB also acknowledges support from the China Postdoctoral Science Foundation through grant 2018M631241. 
LEP acknowledges support of the Russian Foundation for Basic Research (project 18-02-00660). 
DKO acknowledges the support of the Department of Atomic Energy, Government of India, under project No. 12-R\&D-TFR-5.02-0200.
%
%

%
%
\begin{table*}
\setlength{\tabcolsep}{0.07in}
\centering
\caption{Summary of the properties of 53 ATLASGAL dust clumps at 870 $\mu$m. 
These clumps are situated at a distance of $\sim$3.1 kpc. In the table, we have provided ID, Galactic coordinates ({\it l}, {\it b}), 870 $\mu$m integrated flux density ($S_{870}$), radial velocity ($V_\mathrm{lsr}$), clump effective radius ($R_\mathrm{c}$), dust temperature ($T_\mathrm{d}$), 
bolometric luminosity ($L_{\rm bol}$), clump mass ($M_\mathrm{clump}$), H$_\mathrm{2}$ column density ($N(\mathrm H_2)$), 
average volume density ($n_{\mathrm H_2}$), and virial mass ($M_\mathrm{vir}$). Using different symbols, five groups are indicated in Figure~\ref{fig2}a. 
These groups are group1 (c1--c4; up down triangles), group2 (c5--c13; circles), group3 (c14--c22; squares), group4 (c23--c42; triangles), and group5 (c43--c53; stars). The clumps highlighted with dagger have the NH$_{3}$ line-width, and have the ratio $M_\mathrm{clump}$/$M_\mathrm{vir}$ $>$ 1 \citep[see][for more details]{urquhart18}.} 
\label{tab3}
\begin{tabular}{lcccccccccccr}
\hline 
 ID&   {\it l}     &  {\it b}      & $S_{870}$  & $V_\mathrm{lsr}$   &R$_\mathrm{c}$ &$T_\mathrm{d}$ &$L_{\rm bol}$ &$M_\mathrm{clump}$&$N(\mathrm H_2)$ & $n_{\mathrm H_2}$&$M_\mathrm{vir}$ & Association\\  
      & (degree) & (degree)   &  (Jy)      & (km s$^{-1}$)     &     (pc)     & (K)&   (10$^{2}$ $L_\odot$)& ($M_\odot$) &(10$^{22}$ cm$^{-2}$)  &(10$^{3}$ cm$^{-3}$)& ($M_\odot$) &\\  
\hline
\hline 

    c1 &     13.786 &  -0.237 &    1.96 &   38.7 &    0.15 &   12.4   &    0.4  &     224.4  &   2.9   &   229.8  &	  -- 	  &  group1  \\
    c2 &     13.867 &  -0.299 &    2.15 &   39.8 &    0.15 &   15.8   &    1.6  &     162.6  &   1.7   &   166.5  &	  -- 	  &  group1  \\
    c3 &     13.882 &  -0.369 &    1.82 &   39.7 &    0.15 &   11.3   &    0.2  &     247.7  &   2.3   &   253.8  &	  -- 	  &  group1  \\
    c4 &     13.906 &  -0.292 &    2.88 &   39.7 &    0.34 &   12.4   &    0.9  &     329.6  &   2.6   &    29.0  &	  -- 	  &  group1  \\
 \hline   
    c5 &     13.982 &  -0.144 &   16.60 &   40.0 &    0.48 &   32.2   &   74.3  &     459.2  &   1.8   &    14.4  &	  -- 	  &  group2  \\
    c6 &     13.992 &  -0.121 &    7.94 &   38.1 &    0.15 &   33.7   &   20.1  &     207.5  &   0.9   &   212.5  &	  -- 	  &  group2  \\
    c7 &     13.997 &  -0.156 &   17.33 &   40.2 &    0.44 &   27.6   &   30.6  &     586.1  &   1.4   &    23.8  &	  -- 	  &  group2  \\
    c8 &     14.001 &  -0.109 &    5.60 &   37.9 &    0.15 &   33.0   &   10.8  &     150.3  &   0.7   &   154.0  &	  -- 	  &  group2  \\
    c9 &     14.009 &  -0.106 &    4.13 &   40.2 &    0.15 &   33.0   &   19.5  &     110.9  &   0.7   &   113.6  &	  -- 	  &  group2  \\
   c10$\dagger$ &     14.011 &  -0.176 &   11.23 &   41.0 &    0.44 &   20.6   &   14.6  &     570.2  &   4.1   &    23.1  &	  337	  &  group2  \\
   c11 &     14.017 &  -0.161 &    6.76 &   41.2 &    0.30 &   19.6   &    7.3  &     369.0  &   2.4   &    47.2  &	  -- 	  &  group2  \\
   c12 &     14.019 &  -0.134 &   22.12 &   40.6 &    0.87 &   26.9   &  162.2  &     774.5  &   3.3   &     4.1  &	  -- 	  &  group2  \\
   c13 &     14.026 &  -0.202 &    1.18 &   37.1 &    0.15 &   16.2   &    0.8  &      86.5  &   1.2   &    88.6  &	  -- 	  &  group2  \\
 \hline  
   c14 &     14.184 &  -0.227 &    9.23 &   39.7 &    0.61 &   14.6   &   10.5  &     790.7  &   3.6   &    12.0  &	  -- 	  &  group3  \\
   c15 &     14.187 &  -0.151 &   12.47 &   38.1 &    0.46 &   15.7   &    2.3  &     948.4  &   2.1   &    33.7  &	  -- 	  &  group3  \\
   c16 &     14.192 &  -0.166 &   10.48 &   35.6 &    0.40 &   15.9   &    3.6  &     779.8  &   2.0   &    42.1  &	  -- 	  &  group3  \\
   c17$\dagger$ &     14.194 &  -0.194 &   42.86 &   39.4 &    1.12 &   16.1   &   37.2  &    3126.1  &  13.7   &     7.7  &	  1694    &  group3  \\
   c18 &     14.197 &  -0.214 &    1.07 &   39.2 &    0.15 &   12.1   &    0.3  &     127.1  &   5.2   &   130.1  &	  -- 	  &  group3  \\
   c19 &     14.204 &  -0.207 &    4.04 &   39.8 &    0.15 &   13.0   &    2.2  &     422.7  &   5.0   &   432.9  &	  -- 	  &  group3  \\
   c20 &     14.206 &  -0.109 &    1.57 &   35.1 &    0.15 &   29.2   &   13.6  &      48.8  &   0.7   &    49.9  &	  -- 	  &  group3  \\
   c21$\dagger$ &     14.231 &  -0.176 &    8.40 &   37.5 &    0.61 &   15.0   &    4.5  &     687.1  &   3.1   &    10.5  &	  333	  &  group3  \\
   c22 &     14.236 &  -0.166 &    2.90 &   35.8 &    0.21 &   10.1   &    0.2  &     488.7  &   4.2   &   182.4  &	  -- 	  &  group3  \\
 \hline  
   c23$\dagger$ &     14.312 &  -0.189 &    3.45 &   38.8 &    0.27 &   15.4   &    1.4  &     270.4  &   2.4   &    47.5  &	  184	  &  group4  \\
   c24 &     14.319 &  -0.147 &    2.27 &   38.6 &    0.15 &   17.5   &    1.1  &     145.2  &   1.3   &   148.7  &	  -- 	  &  group4  \\
   c25 &     14.364 &  -0.102 &    2.84 &   39.7 &    0.15 &   13.9   &    0.4  &     264.2  &   1.8   &   270.7  &	  -- 	  &  group4  \\
   c26 &     14.397 &  -0.181 &    2.63 &   42.6 &    0.21 &   19.4   &    4.4  &     143.9  &   1.1   &    53.7  &	  -- 	  &  group4  \\
   c27 &     14.406 &  -0.052 &    8.22 &   37.7 &    0.82 &   24.0   &   44.4  &     331.9  &   1.2   &     2.1  &	  -- 	  &  group4  \\
   c28 &     14.414 &  -0.069 &    3.53 &   38.0 &    0.58 &   25.7   &   93.1  &     129.7  &   1.2   &     2.3  &	  -- 	  &  group4  \\
   c29 &     14.419 &  -0.056 &    2.23 &   37.7 &    0.15 &   22.0   &    2.7  &     101.6  &   1.3   &   104.1  &	  -- 	  &  group4  \\
   c30 &     14.427 &  -0.096 &    5.11 &   39.6 &    0.37 &   19.9   &    9.2  &     269.2  &   1.4   &    18.4  &	  -- 	  &  group4  \\
   c31 &     14.431 &  -0.056 &    6.96 &   37.7 &    0.31 &   25.0   &   39.5  &     265.5  &   0.8   &    30.8  &	  -- 	  &  group4  \\
   c32$\dagger$ &     14.449 &  -0.101 &   25.78 &   40.2 &    0.76 &   19.1   &   44.8  &    1445.4  &   6.0   &    11.4  &	  1247    &  group4  \\
   c33 &     14.449 &  -0.161 &    4.17 &   37.9 &    0.15 &   12.6   &    1.0  &     460.3  &   2.7   &   471.4  &	  -- 	  &  group4  \\
   c34 &     14.454 &  -0.142 &    1.51 &   39.5 &    0.15 &   17.8   &    2.0  &      94.2  &   1.8   &    96.5  &	  -- 	  &  group4  \\
   c35 &     14.459 &  -0.192 &    4.47 &   39.5 &    0.15 &   15.9   &    0.7  &     332.7  &   1.7   &   340.7  &	  -- 	  &  group4  \\
   c36 &     14.462 &  -0.109 &    9.18 &   35.9 &    0.40 &   20.1   &   12.9  &     476.4  &   4.9   &    25.7  &	  -- 	  &  group4  \\
   c37$\dagger$ &     14.469 &  -0.084 &   17.96 &   38.1 &    0.72 &   13.2   &    7.3  &    1828.1  &   7.2   &    16.9  &	  662	  &  group4  \\
   c38 &     14.469 &  -0.101 &    3.02 &   36.6 &    0.15 &   17.8   &    3.2  &     188.4  &   4.2   &   192.9  &	  -- 	  &  group4  \\
   c39 &     14.476 &  -0.126 &    4.44 &   40.5 &    0.28 &   16.9   &    2.2  &     299.9  &   4.6   &    47.2  &	  -- 	  &  group4  \\
   c40 &     14.484 &  -0.192 &    9.05 &   38.2 &    0.39 &   13.5   &    1.0  &     885.1  &   2.4   &    51.6  &	  -- 	  &  group4  \\
   c41$\dagger$ &     14.492 &  -0.139 &   29.46 &   39.7 &    1.12 &   16.6   &   34.2  &    2046.4  &   9.0   &     5.0  &	  1837    &  group4  \\
   c42 &     14.499 &  -0.027 &    1.90 &   38.0 &    0.15 &   15.1   &    0.6  &     153.8  &   1.6   &   157.6  &	  -- 	  &  group4  \\
 \hline  
   c43 &     14.592 &  -0.122 &    1.72 &   39.0 &    0.15 &   27.5   &    6.4  &      57.7  &   0.6   &    59.1  &	  -- 	  &  group5  \\
   c44 &     14.622 &  -0.131 &    2.43 &   39.7 &    0.30 &   13.7   &    1.1  &     231.7  &   2.1   &    29.7  &	  -- 	  &  group5  \\
   c45 &     14.644 &  -0.117 &    5.02 &   40.8 &    0.46 &   12.0   &    2.7  &     606.7  &   4.1   &    21.5  &	  -- 	  &  group5  \\
   c46 &     14.652 &  -0.001 &    4.36 &   37.1 &    0.30 &   12.1   &    0.7  &     518.8  &   3.6   &    66.4  &	  -- 	  &  group5  \\
   c47$\dagger$ &     14.686 &  -0.222 &   12.92 &   37.7 &    0.78 &   14.5   &    5.4  &    1119.4  &   3.6   &     8.2  &	  495	  &  group5  \\
   c48 &     14.696 &  -0.137 &    2.59 &   38.4 &    0.25 &   10.6   &    0.3  &     396.3  &   3.1   &    87.7  &	  -- 	  &  group5  \\
   c49$\dagger$ &     14.707 &  -0.156 &    9.20 &   40.6 &    0.66 &   14.3   &    5.0  &     814.7  &   3.7   &     9.8  &	  638	  &  group5  \\
   c50 &     14.711 &  -0.224 &    4.66 &   36.4 &    0.42 &	9.7   &    0.4  &     853.1  &   4.9   &    39.8  &	  -- 	  &  group5  \\
   c51$\dagger$ &     14.726 &  -0.202 &   18.98 &   37.5 &    0.70 &   13.0   &    3.0  &    1981.5  &   3.8   &    20.0  &	  156	  &  group5  \\
   c52 &     14.756 &  -0.174 &    8.96 &   38.9 &    0.24 &   19.3   &    0.7  &     494.3  &   1.1   &   123.6  &	  -- 	  &  group5  \\
   c53 &     14.789 &  -0.197 &    8.53 &   39.2 &    0.49 &	8.5   &    0.5  &    2079.7  &   6.7   &    61.1  &	  -- 	  &  group5  \\
\hline          
\end{tabular}
\end{table*}
%
%
\begin{table*}
\setlength{\tabcolsep}{0.15 in}
\centering
\caption{Physical parameters of selected molecular clumps in the direction of group2, group3, 
and group4 using the C$^{18}$O line data, which are indicated in Figure~\ref{mfg}. 
Column~1 shows the IDs assigned to the molecular clump(s). 
Table also contains central positions ({\it l}, {\it b}), C$^{18}$O clump diameter (D$_{c}$), 
mass derived from C$^{18}$O ($M_\mathrm{mc}$), Full Width Half 
Maximum (C$^{18}$O $\Delta V$), M$_\mathrm{vir}$, and mean number density ($\bar n$)
calculated from $M_\mathrm{mc}$ and D$_c$. Note that the C$^{18}$O spectra toward the clumps g3clm5 and g3clm6 
contain two closely located velocity components, which do not allow to determine their line widths. 
Hence the values of M$_{vir}$ are not computed for the clumps g3clm5 and g3clm6.} 
\label{moltab}
\begin{tabular}{lcccccccccccccr}
\hline 
  ID  &  {\it l} &  {\it b}    &  D$_{c}$   & $M_\mathrm{mc}$    &  $\Delta V$     &M$_\mathrm{vir}$    &$\bar n$  & Association              \\  
      & (degree) & (degree)    &  (pc)      & ($M_\odot$)&  (km s$^{-1}$)  & ($M_\odot$) & (10$^{4}$ cm$^{-3}$)  &  \\  
\hline
\hline 
g2clm1 & 14.015 & -0.138 &  2.22 &  4265  &  3.77 &   3360  &  1.30&  group2	 \\ 
g2clm2 & 14.015 & -0.169 &  1.43 &  3930  &  3.03 &   1380  &  4.46&  group2	 \\ 
g2clm3 & 13.992 & -0.146 &  2.50 &  4930  &  3.02 &   2390  &  1.06&  group2	 \\ 
g3clm1 & 14.185 & -0.228 &  1.39 &  3575  &  2.93 &   1250  &  4.46&  group3	 \\ 
g3clm2 & 14.201 & -0.191 &  1.30 &  3530  &  3.85 &   2040  &  5.27&  group3	 \\ 
g3clm3 & 14.198 & -0.173 &  1.43 &  2200  &  3.57 &   1910  &  2.52 &  group3	 \\ 
g3clm4 & 14.191 & -0.158 &  0.60 &  1210  &  3.40 &    733  &  18.20&  group3	 \\ 
g3clm5 & 14.188 & -0.139 &  1.87 &  1630  &   --  &    --   &  0.84&  group3	 \\ 
g3clm6 & 14.195 & -0.129 &  1.11 &  1470  &   --  &    --   &  3.59&  group3	 \\ 
g3clm7 & 14.220 & -0.125 &  1.27 &  1890  &  4.01 &   2140  &  3.07&  group3	 \\ 
g4clm1 & 14.450 & -0.102 &  1.99 & 16230  &  5.43 &   6130  &  6.92&  group4	 \\ 
\hline          
\end{tabular}
\end{table*}
\begin{table*}
\setlength{\tabcolsep}{0.1in}
\centering
\caption{Physical parameters of extended regions identified in the {\it Herschel} column density map, which are highlighted in Figure~\ref{fig7}a. 
Column~1 lists the IDs given to the extended region. Table also contains 
positions, deconvolved effective radius ($R_\mathrm{c}$), and clump mass ($M_\mathrm{clump}$). }
\label{ftab1}
\begin{tabular}{lcccccccccccr}
\hline 
 ID   & {\it l}        & {\it b}    & $R_\mathrm{c}$    &$M_\mathrm{clump}$  & Association \\  
      & [degree]       & [degree]   &  (pc) & ($M_\odot$)   &  \\  
\hline
\hline 
   h1 &  13.789 &  -0.238  &    0.70   &    880 &  group1 \\
   h2 &  13.870 &  -0.301  &    0.74   &   1040 &  group1 \\
   h3 &  13.910 &  -0.293  &    0.53   &    525 &  group1  \\
 \hline  
   h4 &  14.014 &  -0.176  &    1.68   &   6175 &  group2 \\
 \hline  
   h5 &  14.197 &  -0.195  &    1.51   &   5400 &  group3 \\
   h6 &  14.234 &  -0.176  &    0.60   &    665 &  group3 \\
 \hline  
   h7 &  14.399 &  -0.186  &    0.47   &    420 &  group4 \\
   h8 &  14.494 &  -0.140  &    4.44   &  42975 &  group4 \\
   h9 &  14.532 &  -0.073  &    0.49   &    425 &  group4 \\
 \hline  
  h10 &  14.610 &   0.010  &    1.40   &   4045 &  group5 \\
  h11 &  14.689 &   0.024  &    0.56   &    635 &  group5  \\
  h12 &  14.682 &  -0.045  &    0.43   &    335 &  group5  \\
  h13 &  14.759 &  -0.081  &    0.86   &   1405 &  group5  \\
  h14 &  14.714 &  -0.155  &    2.55   &  13135 &  group5  \\
  h15 &  14.647 &  -0.118  &    0.92   &   1670 &  group5 \\
  h16 &  14.604 &  -0.138  &    0.77   &   1085 &  group5  \\
  h17 &  14.615 &  -0.203  &    0.76   &   1055 &  group5  \\
 \hline  
  h18 &  14.624 &  -0.326  &    0.86   &   1395 &  group4  \\
  h19 &  14.562 &  -0.361  &    0.90   &   1415 &  group4  \\
  h20 &  14.589 &  -0.408  &    0.60   &    710 &  group4 \\
  h21 &  14.692 &  -0.416  &    0.52   &    490 &  group4  \\
\hline          
\end{tabular}
\end{table*}

%
\begin{table*}
\setlength{\tabcolsep}{0.1in}
\centering
\caption{Physical parameters of 17 radio clumps traced in the GMRT 1280 MHz continuum map, which are labeled in Figure~\ref{fig4}d. 
Table contains ID, Galactic coordinates ({\it l}, {\it b}), 
deconvolved effective radius of the H\,{\sc ii} region ($R_\mathrm{HII}$), total flux ($S{_\nu}$), 
Lyman continuum photons ($\log{N_\mathrm{uv}}$), dynamical age ($t_\mathrm{dyn}$), and radio spectral type.} 
\label{gtab2}
\begin{tabular}{lcccccccccccr}
\hline 
  ID  &  {\it l}     &  {\it b}    &  $R_\mathrm{HII}$   & $S{_\nu}$  & $\log{N_\mathrm{uv}}$  &$t_\mathrm{dyn}$  &Spectral Type\\  
      & (degree) & (degree) &  (pc)               &  (mJy)      &  (s$^{-1}$)           & ($\times$ 10$^{3}$ yr)   &     (dwarf main-sequence (V))  \\  
\hline
\hline 
     g1  &  13.999 &  -0.112  &  0.10  &   26  &  46.29 &  1.1  &  B0.5-B1 \\
     g2  &  14.006 &  -0.123  &  0.27  &   238 &  47.25 &  9.8  &  B0-B0.5 \\	
     g3  &  14.007 &  -0.127  &  0.17  &   98  &  46.87 &  4.0  &  B0-B0.5 \\	
     g4  &  14.008 &  -0.130  &  0.21  &   142 &  47.03 &  5.8  &  B0-B0.5 \\  
     g5  &  13.995 &  -0.118  &  0.20  &   121 &  46.96 &  6.3  &  B0-B0.5 \\	
     g6  &  13.997 &  -0.123  &  0.17  &   97  &  46.86 &  4.2  &  B0-B0.5 \\	
     g7  &  13.994 &  -0.121  &  0.15  &   74  &  46.75 &  3.0  &  B0-B0.5 \\	
     g8  &  13.997 &  -0.127  &  0.16  &   77  &  46.76 &  3.7  &  B0-B0.5 \\	
     g9  &  14.000 &  -0.135  &  0.13  &   47  &  46.55 &  2.6  &  B0-B0.5 \\	
    g10  &  14.002 &  -0.136  &  0.14  &   58  &  46.64 &  3.3  &  B0-B0.5 \\	
    g11  &  13.987 &  -0.119  &  0.20  &   133 &  47.00 &  5.8  &  B0-B0.5 \\	
    g12  &  13.987 &  -0.125  &  0.14  &   68  &  46.71 &  2.2  &  B0-B0.5 \\	
    g13  &  13.988 &  -0.127  &  0.14  &   62  &  46.66 &  2.3  &  B0-B0.5 \\	
    g14  &  13.990 &  -0.129  &  0.11  &   35  &  46.42 &  1.5  &  B0.5-B1 \\	
    g15  &  13.990 &  -0.134  &  0.20  &   127 &  46.98 &  5.1  &  B0-B0.5 \\	
    g16  &  13.990 &  -0.140  &  0.11  &   34  &  46.41 &  1.7  &  B0.5-B1 \\	
    g17  &  13.982 &  -0.125  &  0.13  &   46  &  46.54 &  2.3  &  B0-B0.5 \\	   
\hline          
\end{tabular}
\end{table*}

\end{document}